\documentclass[preprint,prd,tightenlines,nofootinbib]{revtex4}

\pdfoutput=1
\usepackage{amsmath,bm}
\usepackage{graphicx}
\usepackage{hyperref} 

\newcommand{\as}{\alpha_{\mathrm{s}}}
\newcommand{\LA}{\mathrm{A}}
\newcommand{\LB}{\mathrm{B}}
\newcommand{\LF}{\mathrm{F}}

\newcommand{\LI}{\mathrm{I}}

\newcommand{\LL}{\mathrm{L}}
\newcommand{\LR}{\mathrm{R}}
\newcommand{\LS}{\mathrm{S}}
\newcommand{\LT}{\mathrm{T}}
\newcommand{\La}{\mathrm{a}}
\newcommand{\Lb}{\mathrm{b}}

\newcommand{\Lf}{\mathrm{f}}

\newcommand{\Lp}{\mathrm{p}}

\newcommand{\LMC}{\mathrm{MC}}

\newcommand{\tildenot}{{\raise.17ex\hbox{$\scriptstyle\mathtt{\sim}$}}} 

\begin{document}

\title{Finding physics signals with shower deconstruction}

\author{Davison E. Soper}
\affiliation{
Institute of Theoretical Science\\
University of Oregon\\
Eugene, OR  97403-5203, USA\\
}

\author{Michael Spannowsky}
\affiliation{
Institute of Theoretical Science\\
University of Oregon\\
Eugene, OR  97403-5203, USA\\
}

\begin{abstract}
We introduce shower deconstruction, a method to look for new physics in a hadronic environment. The method aims to be a full information approach using small jets. It assigns to each event a number $\chi$ that is an estimate of the ratio of the probability for a signal process to produce that event to the probability for a background process to produce that event. The analytic functions we derive to calculate these probabilities mimic what full event generators like {\sc Pythia} or {\sc Herwig} do and can be depicted in a diagrammatic way. As an example, we apply this method to a boosted Higgs boson produced in association with a $Z$-boson and show that this method can be useful to discriminate this signal from the $Z$+jets background.

\end{abstract}

\date{2 August 2011}

\pacs{}
\maketitle
\section{Introduction}

A central problem for data analysis at the Large Hadron Collider (LHC) is to find the signal for the production of a new heavy particle or particles against a background of jets produced by standard model processes that do not involve the sought heavy particle. Examples include searches for supersymmetric partners of the quarks and gluons and searches for the Higgs boson. While such searches focus on leptonic final states, most of the sought new physics resonances have a large branching ratio to hadrons. Thus, it is of great importance to be able to disentangle hadronically decaying particles with masses around the electroweak scale from large QCD backgrounds. 

The decay products of a new very heavy particle will appear in the detector as one or more jets. There may also be jets from initial state radiation. The jets will contain subjets. In this paper, we call the subjets {\em microjets}. They are defined with a standard jet algorithm but with a small effective cone size $R$.  The pattern of microjets in events arising from the new particle decay will differ from the pattern of microjets in background events that do not involve new particles. One can take advantage of this difference to separate signal from background.

In this paper, we propose a method for separating signal from background by analyzing the distribution of the microjets. This method has the potential to be effective in quite general circumstances. However, for a first application, we choose a process in which we are looking at the microjets contained in a larger jet that results from the decay of a heavy particle with large transverse momentum, that is a highly boosted heavy particle.

There are several methods already available for the analysis of the structure of the microjets produced by the decay of a highly boosted heavy particle. Two of these methods, trimming \cite{Krohn:2009th} and pruning \cite{Ellis:2009su,Ellis:2009me} can be characterized as generic in that they have the potential to discover new physics signals even if one does not have in mind a particular new physics scenario.  Other methods, including the one proposed here, are adapted to searches for particular new physics signals. These include mass drop with filtering and b-quark tagging \cite{Butterworth:2008iy}, the matrix element method \cite{Kondo:1988yd,Kondo:1991dw,Fiedler:2010sg,Alwall:2010cq}, and the template overlap method \cite{Almeida:2010pa}. These last two methods bear some resemblance to the method proposed in this paper. One can also combine methods \cite{Soper:2010xk}. For further applications see Refs.~\cite{Butterworth:2002tt, Butterworth:2009qa, Seymour:1993mx, Thaler:2008ju, Kaplan:2008ie,Plehn:2009rk, Chen:2010wk,Falkowski:2010hi, Kribs:2009yh, Kribs:2010hp, Plehn:2010st, Bhattacherjee:2010za, Hackstein:2010wk, Englert:2010ud, Katz:2010mr, Almeida:2008yp, Thaler:2010tr, Kim:2010uj, Kribs:2010ii,Fan:2011jc,Plehn:2011tf} and for a review see Ref.~\cite{Abdesselam:2010pt}.

The example that we consider in this paper is the production of a Higgs boson in association with a high transverse momentum $Z$-boson, where the $Z$-boson decays into $e^+ + e^-$ or $\mu^+ + \mu^-$ and the Higgs boson decays into $b + \bar b$. This example was analyzed in Ref.~\cite{Butterworth:2008iy}. Since the Higgs boson recoils against a high transverse momentum $Z$-boson, the Higgs boson has a large transverse momentum and is easier to find than if it had low transverse momentum. Nevertheless, there is a large background to this process from standard model processes that do not involve the Higgs boson, so some ingenuity is required to separate the signal from the background. 

The idea of this paper is to define an observable $\chi$ that is a function of the observed configuration of the final state microjets in an event and distinguishes between a sought signal and the background. To do that, we define $\chi$ as the ratio of the probability that the microjet configuration observed would arise in a signal event to the probability that it would arise in a background event. We use a parton shower algorithm for this purpose. However, our parton shower algorithm is massively simplified compared to {\sc Pythia} \cite{Pythia} or {\sc Herwig} \cite{Herwig} in order that we can compute the probability for a given microjet configuration analytically. We call the method proposed here shower deconstruction.

\section{Overview and event selection}

As stated in the introduction, the idea of this paper is to define an observable $\chi$ that is a function of the configuration of the final state in an event and distinguishes between a sought signal and the background. The method that we propose is quite general, but in order to explain it with reasonable clarity, we need to consider a specific process.  Our choice of process is guided by the desire to have a case that is relatively simple to explain. The example that we use is the search for the Higgs boson using the process $p+p \to H + Z + X$ where the $Z$-boson decays to $\mu^+ + \mu^-$ (or $e^+ + e^-$) while the Higgs boson $H$ decays to $b + \bar b$.  We try to separate this from the background process $p+p \to {\it jets} + Z + X$ \cite{Butterworth:2008iy}. 

\subsection{Event selection}
\label{sec:EventSelection}

We simulate an analysis of data by using events generated by {\sc Pythia} \cite{Pythia}. In order to make the Higgs boson easier to find, we demand that the $Z$-boson against which it recoils has a large transverse momentum.  Specifically, we select events consistent with a leptonically decaying $Z$-boson for which the leptons are central ($|y_l| <2.5$) and fairly hard ($p_{T,l} > 15~\rm{GeV} $). The invariant mass of the leptons is required to match the $Z$-boson mass,
\begin{equation}
|m_{l^+l^-} - m_Z| < 10~\rm{GeV}
\;\;.
\end{equation}
The reconstructed $Z$-boson is required to be highly boosted in the transverse plane,
\begin{equation}
\label{eq:pTcut}
p_{T,l^+l^-} > p_{T,{\rm min}} \equiv 200~\rm{GeV}
\;\;.
\end{equation}

We next combine final state hadrons in simulated detector cells of size $0.1\times 0.1$ and adjust the absolute value of the momentum in each cell so that the four-momentum is massless. We remove cells with energy less than $0.5\ {\rm GeV}$. We then use these cells as input to  the anti-$k_T$ jet-finding algorithm \cite{antiKT} with a large effective cone size, $R_\LF = 1.2$. For the recombination of the jet constituents we use {\sc Fastjet} \cite{Cacciari:2005hq}. We find the jet with the highest transverse momentum of all such jets in the event and require its transverse momentum to be larger than $p_{T,{\rm min}}$. This is the ``fat jet.''

Those selection cuts force the Higgs boson to recoil against the $Z$-boson with a large transverse momentum, so that the decay products of the Higgs boson are fairly well collimated. 

We denote the cross section for signal events that pass these cuts by $\sigma_\LMC(\LS)$ and denote the cross section for background events that pass these cuts by $\sigma_\LMC(\LB)$. With some help from next-to-leading order calculations, we estimate \footnote{We generate events for $Z+\it{jet}\rightarrow l^+l^-+\it{jet}$ and $HZ\rightarrow b \bar{b}~l^+l^-$ using {\sc Pythia} in a configuration with large transverse momentum and normalize the cross section to the one obtained from {\sc MCFM} \cite{MCFM} with the same cuts. Then we calculate the cross section after selection cuts based on the number of events that pass the selection cuts.}
\begin{equation}
\begin{split}
\label{eq:sigmatotSB}
\sigma_\LMC(\LS) ={}& 1.57\ {\rm fb}
\;\;,
\\
\sigma_\LMC(\LB) ={}& 2613\ {\rm fb}
\;\;,
\\
\frac{\sigma_\LMC(\LS)}{\sigma_\LMC(\LB)} ={}& \frac{1}{1664}
\;\;.
\end{split}
\end{equation}
Our analysis makes use of events generated by a Monte Carlo event generator that we use and regard as an accurate representation of nature. We renormalize the event generator cross sections by constant factors for signal and background calculations so as to match the cross sections given in Eq.~(\ref{eq:sigmatotSB}). We will generally use ``MC'' subscripts to denote quantities calculated by a Monte Carlo event generator supplemented by some  next-to-leading order information. As noted above, we use {\sc Pythia} in our calculations; in Sec.~\ref{sec:results}, we also present results using {\sc Herwig}. 

\subsection{Variables describing the final state}
\label{sec:FinalStateVariables}

In principle, the final state could be described by the momenta and flavors of all final state particles. However, we simplify this. First, we select events and use the anti-$k_T$ algorithm to define the ``fat jet'' that recoils against the $Z$-boson, as described above.

We use the $k_T$ jet-finding algorithm \cite{kTjets} to group the fat jet into subjets, which we call microjets. We choose the effective cone size in the $k_T$ jet-finding algorithm to be $R = 0.15$. This size is chosen to correspond roughly to the angular resolution of calorimeter topological clusters in the ATLAS experiment and to be a little larger than the ALTAS calorimeter angular resolution of about 0.1 \cite{AtlasAngularResolution}.  We do not want any of the microjets to be exactly massless, so we add $0.1\ {\rm GeV}$ to the energy of each microjet. 

Typically, the number of microjets found is between six and ten, but a few events have even more microjets. The computational time needed to analyze an event increases quite quickly with the number of microjets. Accordingly, we choose a number $N_{\rm max}$ with default value $N_{\rm max}=7$ and discard the lowest transverse momentum microjets if there are more than $N_{\rm max}$ microjets, keeping the $N_{\rm max}$ microjets that have the highest transverse momenta.  In fact, we find that the lowest transverse momentum microjets carry little useful information: we have varied $N_{\rm max}$ between 5 and 9 and find that the statistical significance of the results that we obtain, as discussed in Sec.~\ref{sec:conclusions}, increases only slowly with $N_{\rm max}$.

The microjets found by this procedure are described, in part, by their momenta $\{p\}_N = \{p_1, \dots, p_N\}$, with $p_i^2 > 0$.

For some microjets $j$, we also provide a $b$-quark tag, $t_j$. To qualify for a tag, the microjet must be among the three microjets in the event with the highest $p_T$ values and it must have $p_T > p_T^{\rm tag}$, where our default value is $p_T^{\rm tag} = 15\ {\rm GeV}$. For microjets $j$ that do not qualify for a tag we set $t_j = {\tt none}$. In the simplest implementation, one would take $t_j = {\rm T}$ if microjet $j$ contains a $b$ or $\bar  b$ quark and otherwise define $t_j = {\rm F}$.  We simulate $b$-tagging of microjets in experiment by using more realistic $b$-tagging for {\sc Pythia} events:
\begin{quote}

$\bullet$ If any hadron in microjet $j$ contains a $b$ or $\bar b$ quark, then we set $t_{j} = \LT$ with a probability $P(\LT|b)$ and $t_{j} = \LF$ with a probability $1-P(\LT|b)$. 

$\bullet$ If no hadron in microjet $j$ contains a $b$ or $\overline b$ quark, then we set that $t_{j} = \LT$ with a probability $P(\LT|\tildenot b)$ and $t_{j} = \LF$ with a probability $1-P(\LT|\tildenot b)$. 

\end{quote}
Our default value for the $b$-tagging efficiency is $P(\LT|b) = 0.6$ while our default value for the mistag probability is $P(\LT|\tildenot b) = 0.02$ \cite{btags}.

This procedure of defining microjets within the fat jet gives a somewhat ``coarse grained'' description of the part of the event that is of interest: the momenta and b-quark tags, $\{p,t\}_N = \{p_1, t_1; \dots; p_N,t_N\}$, of the microjets. 

\subsection{Probabilities according to Monte Carlo event generator}

We denote by $P_\LMC(\{p,t\}_N|\LS)$ the probability that a signal event has a microjet configuration $\{p,t\}_N$, as determined by the Monte Carlo event generator that we use and regard as an accurate representation of nature:\footnote{Here the differential $dp_j$ for each microjet $j$ can just mean $d^4p_j$.}
\begin{equation}
P_\LMC(\{p,t\}_N|\LS) = \frac{1}{\sigma_\LMC(\LS)}\,
\frac{d\sigma_\LMC(\LS)}{d\{p,t\}_N}
\;\;.
\end{equation}
Similarly, we let the probability that a background event has a microjet configuration $\{p,t\}_N$ be
\begin{equation}
P_\LMC(\{p,t\}_N|\LB) = \frac{1}{\sigma_\LMC(\LB)}\,
\frac{d\sigma_\LMC(\LB)}{d\{p,t\}_N}
\;\;.
\end{equation}

We now seek an observable that does a good job of distinguishing signal events from background events. Our sought observable is to be a function $\chi(\{p,t\}_N)$ of the microjet configuration. It will also be a function of the parameters of the standard model, especially the mass $m_{H}$ of the Higgs boson.

As a preliminary step, we define a quantity $\chi_\LMC(\{p,t\}_N)$ by 
\begin{equation}
\chi_\LMC(\{p,t\}_N) = 
\frac{P_\LMC(\{p,t\}_N|\LS)}{P_\LMC(\{p,t\}_N|\LB)}
\;\;.
\end{equation}
We would like to use $\chi_\LMC(\{p,t\}_N)$ as our observable. In fact, if one considers that the Monte Carlo event generator is accurate and if one could construct $\chi_\LMC$ as a function of $\{p,t\}_N$, then this could be considered to be the ideal observable. 

Why might one consider $\chi_\LMC$ to be an ideal observable? To see this in the simplest context, let us suppose that we want to examine data using a cut: we accept events if $C(\{p,t\}_N) > 0$, where $C(\{p,t\}_N)$ is some function that we are at liberty to make up. The signal and background cross sections with this cut are
\begin{equation}
\begin{split}
\sigma_C(\LS) ={}& \int\! d\{p,t\}_N\ \Theta(C(\{p,t\}_N))\, \frac{d\sigma_\LMC(\LS)}{d\{p,t\}_N}
\;\;,
\\
\sigma_C(\LB) ={}& \int\! d\{p,t\}_N\ \Theta(C(\{p,t\}_N))\, \frac{d\sigma_\LMC(\LB)}{d\{p,t\}_N}
\;\;.
\end{split}
\end{equation}
Choose a value $\sigma_C(\LS)$ that we want for the signal cross section and require that the cut produce this value of signal cross section. With this constraint on the signal cross section, we will have the best statistical significance for a measurement if we make $\sigma_C(\LB)$ as small as possible. Thus we seek to choose the cut so as to minimize $\sigma_C(\LB)$ with $\sigma_C(\LS)$ held constant. The solution to this problem is to choose $C(\{p,t\}_N)$ such the surface $C(\{p,t\}_N) = 0$ is a surface of constant $\chi_\LMC(\{p,t\}_N)$. That is, we should measure the cross section inside a cut defined by
\begin{equation}
C(\{p,t\}_N) = \chi_\LMC(\{p,t\}_N) - \chi_0
\end{equation}
for some $\chi_0$. If we make any small adjustment to this by removing an infinitesimal region with $\chi_\LMC(\{p,t\}_N) > \chi_0$ from the cut and adding a region having the same signal cross section but with $\chi_\LMC(\{p,t\}_N) < \chi_0$, we raise the total background cross section within the cut while keeping the signal cross section the same. Thus using contours of $\chi_\LMC(\{p,t\}_N)$ to define our cut is the best that we can do. 

What value of $\chi_0$ should one choose? For a simple optimized cut based analysis with a given amount of integrated luminosity, one would choose $\chi_0$ so as to maximize the ratio of the expected number of signal events to the square root of the expected number of background events. We discuss this further in Sec.~\ref{sec:results}.

Instead of using an optimized cut on $\chi_\LMC$ to separate signal from background, one could imagine using a log likelihood ratio constructed from $\chi_\LMC$. We do not discuss that method in this paper.

Now we must face the fact that to construct $\chi_\LMC(\{p,t\}_N)$, we would need two things: the differential cross section to find microjets $\{p,t\}_N$ in background events and then the differential cross section to find microjets $\{p,t\}_N$ in signal events. In each case, we would consider this differential cross section in a parton shower approximation to the full theory. Unfortunately for us, a parton shower produces $d\sigma_\LMC(\LS)/d\{p,t\}_N$ and $d\sigma_\LMC(\LB)/d\{p,t\}_N$ by producing Monte Carlo events at random according to these distributions. If we have 7 microjets described by 4 momentum variables each and we divide each of these 28 variables into 10 bins, then we have approximately $10^{28}/7! \approx 10^{24}$ total bins (accounting for the interchange symmetry among the 7 microjets). The parton shower Monte Carlo event generator will fill these bins with events, but it will be a long time before we have of order 100 counts per bin in order to estimate $d\sigma_\LMC(\LS)/d\{p,t\}_N$ and $d\sigma_\LMC(\LB)/d\{p,t\}_N$  at each bin center. Thus it is not practical to calculate $\chi_\LMC(\{p,t\}_N)$ numerically by generating Monte Carlo events. It is also not practical to calculate $\chi_\LMC(\{p,t\}_N)$ analytically using the shower algorithms in \textsc{Pythia} or \textsc{Herwig}. These programs are very complicated, so that we have no hope of finding $P_\LMC(\{p,t\}_N|\LS)$ and $P_\LMC(\{p,t\}_N|\LB)$ for either of them.

\subsection{Probabilities according to simplified shower}
\label{sec:probabilities}

What we need is an observable $\chi(\{p,t\}_N)$ that is an approximation to $\chi_\LMC(\{p,t\}_N)$ such that we can calculate $\chi(\{p,t\}_N)$ analytically for any given $\{p,t\}_N$. For this purpose, we define a simple, approximate shower algorithm, which we will call the simplified shower algorithm. We let $P(\{p,t\}_N|\LS)$ and $P(\{p,t\}_N|\LB)$ be the probabilities to produce the microjet configuration $\{p,t\}_N$ in, respectively, signal and background events according to the simplified shower algorithm. Define
\begin{equation}
\label{eq:chidef}
\chi(\{p,t\}_N) = 
\frac{P(\{p,t\}_N|\LS)}{P(\{p,t\}_N|\LB)}
\;\;.
\end{equation}
This function, $\chi(\{p,t\}_N)$ without the ``MC'' subscript, is the observable that we use. We may call the calculation of $\chi(\{p,t\}_N)$ shower deconstruction.

The parton state with $N$ microjets is a possible intermediate state in a parton shower. We seek to determine the probability that this intermediate state with parameters $\{p,t\}_N$ is generated. We try to build enough into the simpler shower to provide a reasonable approximation to QCD and the rest of the standard model. Furthermore, we can define the shower so that the deconstruction is as simple as we can make it, even if that means that the corresponding shower algorithm is not so practical as an event generator. For instance, an implementation of the simplified shower algorithm as an event generator might generate weighted events in a way that makes unweighting the events costly in computer time. Additionally, probability conservation might be only approximate, so that the generated weights for different outcomes do not sum exactly to one. No matter: we are not going to use the simplified shower algorithm to generate events anyway. Additionally, we can ignore any factors in $P(\{p,t\}_N|\LS)$ and $P(\{p,t\}_N|\LB)$ that are common between them for each $\{p,t\}_N$ since such factors cancel in $\chi$.

\begin{figure}
\centerline{\includegraphics[width=8.0cm]{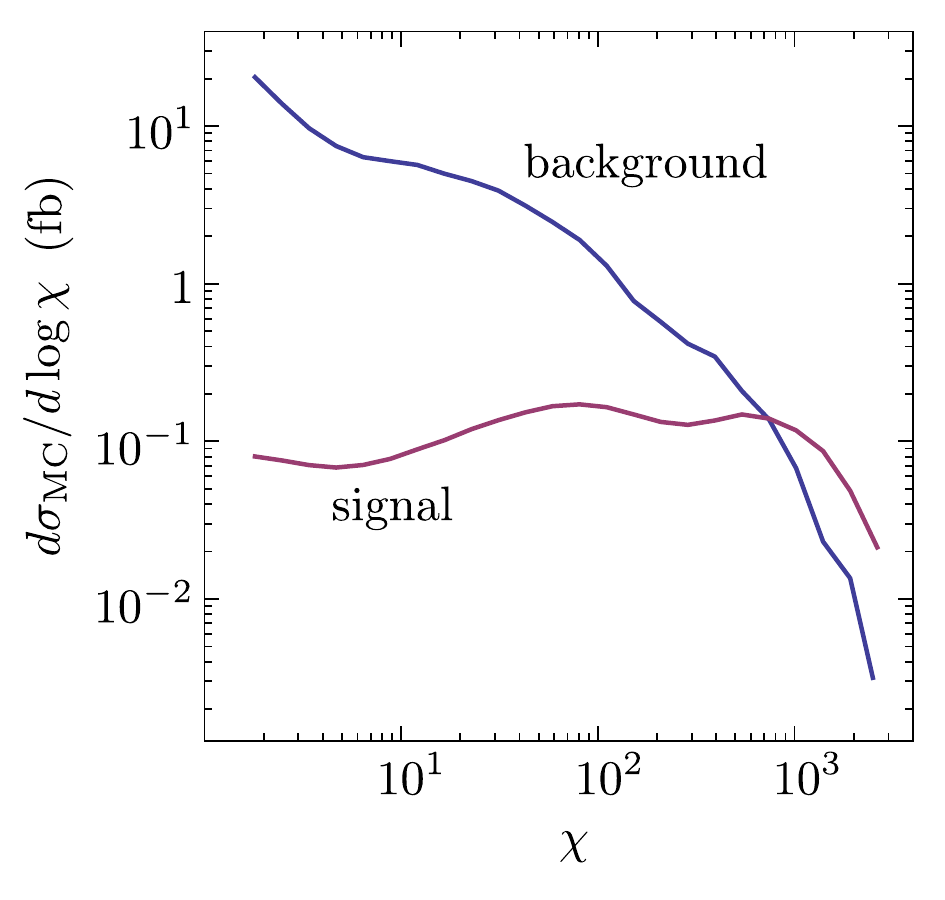}}
\caption{
$d\sigma_\LMC(\LB)/ d\log\chi$ for background events (upper curve) and $d\sigma_\LMC(\LS)/ d\log\chi$ for signal events (lower curve) for samples of signal and background events generated by \textsc{Pythia}. We use the cuts described in Sec.~\ref{sec:EventSelection}.
}
\label{fig:SandBvschi}
\end{figure}

Our construction will be far from perfect, and it can be useful even if it is not perfect. We will use \textsc{Pythia} to measure the cross section $d\sigma_\LMC(\LS)/ d\log\chi$ to have signal events with a given value of $\chi$ and the corresponding cross section $d\sigma_\LMC(\LB)/ d\log\chi$ to have background events with this value of $\chi$. In Fig.~\ref{fig:SandBvschi}, we show these two functions for the simplified shower as defined in the following sections. In this illustration, we see that increasing $\chi$ favors signal compared to background.

There is another way to present the results in Fig.~\ref{fig:SandBvschi} that is more informative. Let us define integrated signal and background cross sections above a cut:
\begin{equation}
\begin{split}
\label{eq:sandbdef}
s(\chi) ={}& \int_{\chi}^\infty\!d\bar\chi\ 
\frac{d\sigma_\LMC(\LS)}{d\bar\chi}
\;\;,
\\
b(\chi) ={}& \int_{\chi}^\infty\!d\bar\chi\ 
\frac{d\sigma_\LMC(\LB)}{d\bar\chi}
\;\;.
\end{split}
\end{equation}
It is useful to use $s$ in plots as the independent variable. With this definition, $s$ runs from 0 to $\sigma_\LMC(\LS)$ and $s = 0$ corresponds to $\chi = \infty$. We can then examine the ratio of signal to background cross sections, $s/b$, considered as a function of $s$.

\begin{figure}
\centerline{\includegraphics[width=8.0cm]{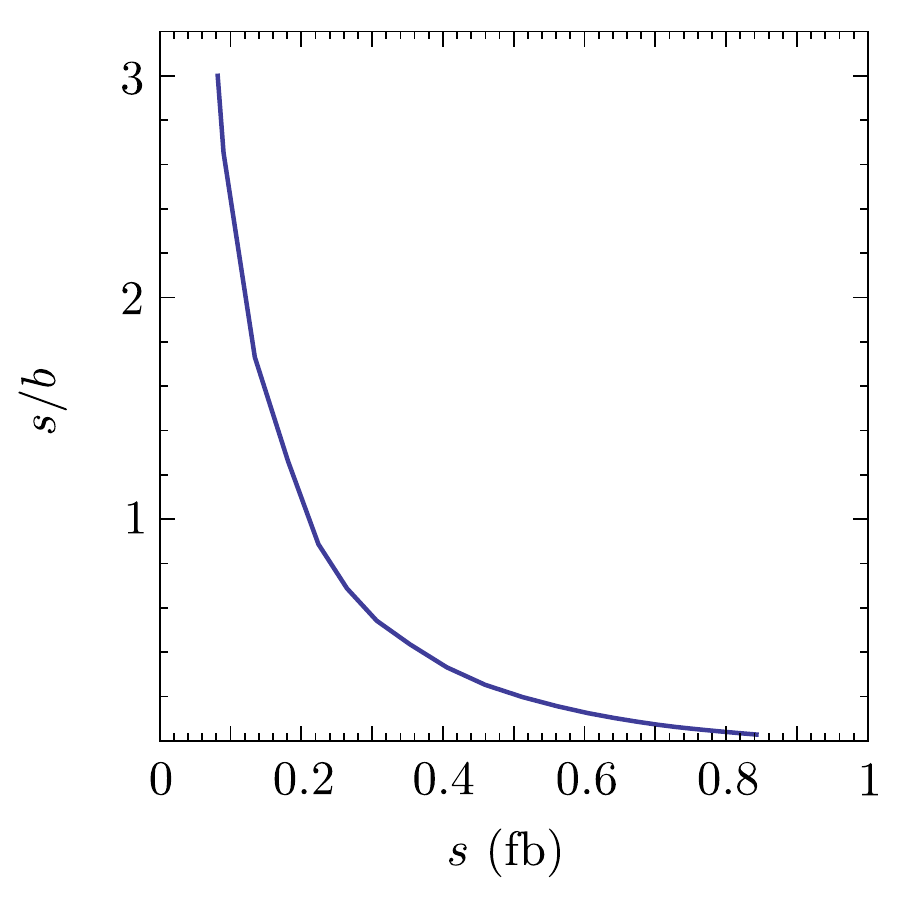}}
\caption{
Plot of $s/b$ versus $s$, where $s$ and $b$ are defined in Eq.~(\ref{eq:sandbdef}). We use samples of signal and background events generated by \textsc{Pythia} as in Fig.~\ref{fig:SandBvschi}. 
}
\label{fig:Rvss}
\end{figure}

In Fig.~\ref{fig:Rvss}, we display the information in Fig.~\ref{fig:SandBvschi} as a plot of $s/b$ versus $s$.  We have used here the $\chi(\{p,t\}_N)$ from our simplified shower algorithm. If we could somehow use $\chi_\LMC(\{p,t\}_N)$, based on the same Monte Carlo event generator that we used to generate events, then we would obtain a curve for $s/b$ versus $s$ that is everywhere higher. No algorithm could produce a curve above this limiting curve, but we have no way of determining the limiting curve.

We see in Fig.~\ref{fig:Rvss} that $s/b$ is small for large $s$ but that there is a region of $s$ in which $s/b$ is not too small. This is what one hopes to accomplish with shower deconstruction. We will return in Sec.~\ref{sec:results} to a discussion of numerical results.

In the following sections, we describe how shower deconstruction works. Conceptually, it is very simple. However, there are quite a few ingredients. That is because we seek to approximate the probability that a parton shower will give a certain set of microjets and there are quite a few ingredients in a parton shower. The simplified parton shower that we describe in the following sections is modeled on the general parton shower algorithm described in Ref.~\cite{NSI} and, in particular, on its leading color, spin-averaged version \cite{NSII}. It is basically a virtuality ordered shower, although we modify the evolution variable in Refs.~\cite{NSI,NSII} to be virtuality/energy instead of just virtuality. This shower is a partitioned dipole shower, and we choose a dipole partitioning function from Ref.~\cite{NSIII}.

A shower algorithm in which one can calculate the probability to produce a given parton configuration has been proposed in Ref.~\cite{Bauer:2008qj}. The aims of this algorithm are rather different from ours in that the algorithm of Ref.~\cite{Bauer:2008qj} is designed to be practical as an event generator. Accordingly, the methods used are rather different from ours.

\section{Organization of shower deconstruction}

In this section, we explain the overall organization of shower deconstruction, beginning with the concept of a shower history.

\subsection{Shower histories}
\label{sec:histories}

\begin{figure}
\centerline{\includegraphics[width=14.0cm]{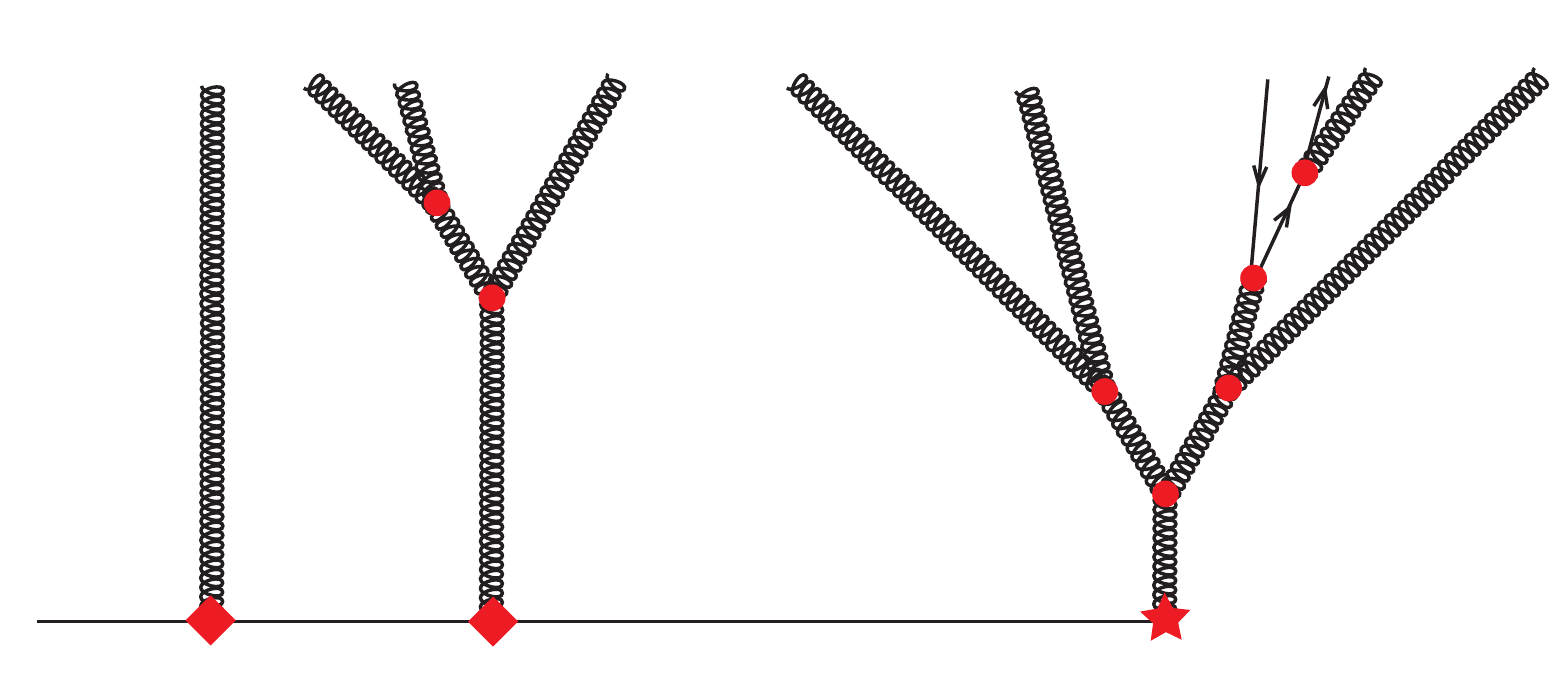}}
\caption{A shower history for a background event. The ``star'' vertex represents the production of a high $p_T$ parton from the hard interaction. The ``diamond'' vertices represent production of partons by initial state radiation. Each parton can split into two daughter partons at a shower vertex, represented by a small circle. In this background event, one of the gluons splits into a light $q$-$\bar q$ pair.}
\label{fig:HistoryBackground}
\end{figure}

\begin{figure}
\centerline{\includegraphics[width=14.0cm]{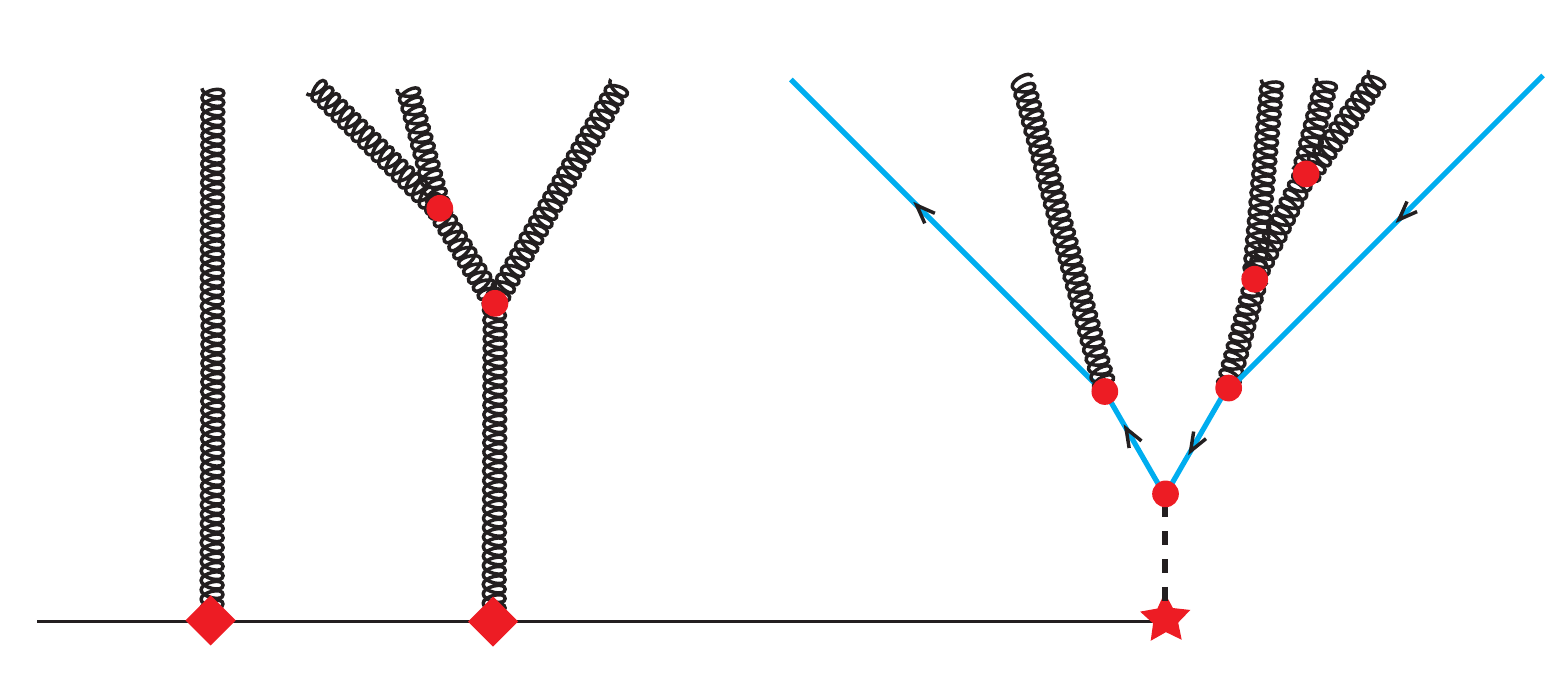}}
\caption{A shower history for a signal event. The dashed line is the Higgs boson, produced in the hard interaction. It decays into a $b$-quark and a $\bar b$-quark, which carry arrows representing the flow of $b$-flavor. The QCD shower splitting of a $b$-quark is to a $b$-quark plus a gluon. In this event, one of the gluons splits into further gluons.}
\label{fig:HistorySignal}
\end{figure}

In general, a shower history $h$ is a tree Feynman diagram showing how $N$ final state partons (the microjets) could have evolved starting with a hard scattering process for signal or background events. In our application, we simplify quite a lot. First, we look not at the whole event, but only at the microjets that make up the fat jet. For background events, we assume that the microjets came from a parton shower induced by a high $p_T$ parton plus parton showers starting from initial state radiation (including radiation from the underlying event), as illustrated in Fig.~\ref{fig:HistoryBackground}.
For signal events, we assume that the microjets came from the decay products of a Higgs boson (through $H \to b + \bar b$) plus parton showers starting from initial state radiation, as illustrated in Fig.~\ref{fig:HistorySignal}. 
 
Each parton in the shower history carries a flavor label $f_i$. We make some  simplifications in the flavor structure of the simplified shower.

\begin{enumerate}

\item For shower histories corresponding to signal events, we have a Higgs boson intermediate state. That is, we have a parton with flavor $f_i = H$.

\item  The Higgs boson decays into a $b$-quark and a $\bar b$-quark, so we need flavors $f_i = b$ and $f_i = \bar b$.

\item A $b$- or $\bar b$-quark can emit a gluon, so we have partons in our shower histories with flavor $f_i = g$.

\item A gluon can split to a $b$-quark and a $\bar b$-quark.

\item A gluon can also split to a light quark and a light antiquark, so we have partons in our shower with flavors $f_i = q$ and $f_i = \bar q$. We do not distinguish whether the light quark pairs  are $({\mathrm u}, \bar {\mathrm u})$, $({\mathrm d}, \bar {\mathrm d})$, $({\mathrm s}, \bar {\mathrm s})$, or $({\mathrm c}, \bar {\mathrm c})$. Instead, we simply multiply the emission probability for one flavor of light quark by $n_\Lf - 1 = 4$, where $n_\Lf = 5$ is the number of quark flavors including the $b$ quark.

\item As an approximation, we treat the initial hard parton in a background event as being a gluon. Similarly, we treat partons radiated from the incoming initial state partons as being gluons.

\end{enumerate}

\begin{figure}
\centerline{\includegraphics[width=14.0cm]{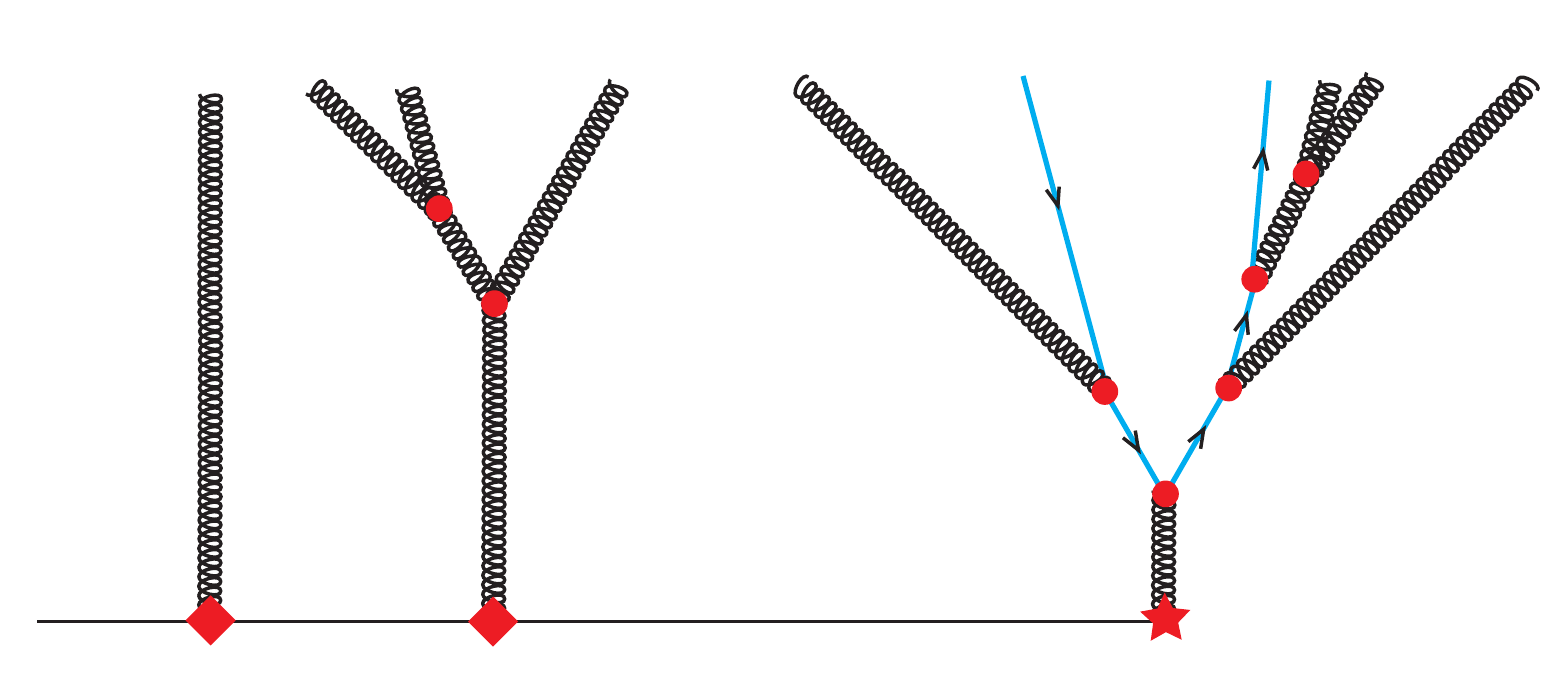}}
\caption{A shower history for a background event in which a high $p_T$ gluon splits to a $b + \bar b$ pair. The QCD shower splitting of a $b$-quark is to a $b$-quark plus a gluon. The $b$ and $\bar b$ quarks radiate gluons and one of the gluons splits into two gluons.}
\label{fig:HistoryBackgroundgbb}
\end{figure}

A shower history in which a gluon splits into a $b$-$\bar b$ pair is illustrated in Fig.~\ref{fig:HistoryBackgroundgbb}.

The probabilities $P(\{p,t\}_N|\LB)$ and $P(\{p,t\}_N|\LS)$ in our shower model will consist of a sum of partial probabilities corresponding to different shower histories. In the following sections, we assume that we have picked a shower history $h$ and we seek to construct the probability $P(\{p,t\}_N|\LB,h)$ or $P(\{p,t\}_N|\LS,h)$ corresponding to that shower history.

We will return in Sec.~\ref{ConstructingHistories} to the question of how to construct the shower histories in a reasonably efficient fashion. First, though, we need to define the factors corresponding to the vertices and propagators in our shower history diagrams. We begin with a description of the color flow.

\subsection{Color connections}

We work in the standard leading color approximation and will need to keep track of color connections. 

Consider a final state splitting in which a gluon labeled $J$ splits into two daughter gluons. Let the label of the daughter that carries the $\overline {\bm 3}$ color of the mother parton $J$ be $A$. We draw this daughter parton on the left in our diagrams. Let the label of the daughter parton that carries the ${\bm 3}$ color of parton $J$ be $B$. We draw this daughter parton on the right in our diagrams. We track the angle variables of two color connected partner partons to parton $J$. Parton $k{(J)}_{\LL}$ carries the $\bf 3$ color that is connected to the $\overline {\bm 3}$ color line of parton $J$. Parton $k{(J)}_{\LR}$ carries the $\overline {\bm 3}$ color that is connected to the $\bf 3$ color line of parton $J$. The labels $k{(J)}_{\LL}$ and $k{(J)}_{\LR}$ specify lines in the shower history diagram, not necessarily final microjets. Given the labels of the color connected partners to the mother parton $J$, we assign the color connected partners of the daughter partons. The two daughter partons are color connected partners of each other and each inherits one of the color connected partners of the mother. That is
\begin{equation}
k{(A)}_\LL = k{(J)}_\LL, \qquad k{(A)}_\LR = B
\;\;,
\end{equation}
and
\begin{equation}
k{(B)}_\LL = A,  \qquad k{(B)}_\LR = k{(J)}_\LR
\;\;.
\end{equation}

If parton $J$ is a quark, then it has a color connected partner $k(J)_\LR$ that carries the $\overline {\bm 3}$ color connected to the quark's $\bm 3$ color. There is no $k(J)_\LL$ partner. The quark can split into daughter quark $A$ and a daughter gluon $B$, which we draw on the right because it carries the $\bm 3$ color of the mother quark. The color connected partners of the daughter partons are then
\begin{equation}
k{(A)}_\LR = B
\;\;,
\end{equation}
and
\begin{equation}
k{(B)}_\LL = A,  \qquad k{(B)}_\LR = k{(J)}_\LR
\;\;.
\end{equation}
Similarly, if parton $J$ is an antiquark, then it has a color connected partner $k(J)_\LL$ that carries the ${\bm 3}$ color connected to the antiquark's $\overline {\bm 3}$ color. There is no $k(J)_\LR$ partner. The antiquark can split into daughter antiquark $B$ and a daughter gluon $A$, which we draw on the left because it carries the $\overline {\bm 3}$ color of the mother antiquark. The color connected partners of the daughter partons are then
\begin{equation}
k{(A)}_\LL = k{(J)}_\LL,  \qquad k{(A)}_\LR = B
\;\;,
\end{equation}
and
\begin{equation}
k{(B)}_\LL = A
\;\;.
\end{equation}

Consider a final state splitting in which a gluon with label $J$ splits into  $q + \bar q$ ( or $b + \bar b$). Let the label of the daughter antiquark be A; we draw it to the left because it carries the $\overline {\bm 3}$ color of the mother parton $J$. Let the label of the daughter quark be B; we draw it to the right because it carries the ${\bm 3}$ color of the mother parton.  The color connected partners of the daughter partons are
\begin{equation}
k{(A)}_\LL = k{(J)}_\LL 
,  \qquad k{(B)}_\LR = k{(J)}_\LR
\;\;.
\end{equation}

Finally, consider the decay of a Higgs boson, labelled $J$, into  $b + \bar b$. Since the Higgs boson is a color singlet, the $b$ and $\bar b$ quarks are each other's color connected partners. We draw the $b$-quark on the left and call its label $A$, while we draw the $\bar b$-quark on the right and call its label $B$. The color connected partners of the daughter partons are
\begin{equation}
k{(A)}_\LR = B
,  \qquad 
k{(B)}_\LL = A
\;\;.
\end{equation}

These procedures define color connections recursively. To start the recursion the initial hard parton in a background event has undefined color connected partners: $k{(J)}_\LL = k{(J)}_\LR = {\tt undefined}$. If we knew the complete Feynman diagram representing a shower history, then all color connected partners would be defined, but we know about only partons that are part of the fat jet, so we have an incomplete shower history. The true color partners of the initial hard parton could be partons that are not in the fat jet, or they could be partons from initial state radiation. Because we do not know the true color connections, we leave them undefined. Similarly, partons created as initial state radiation have {\tt undefined} color connections in our approximation.  As the shower progresses, the {\tt undefined} color connections are inherited, but most partons later in the shower have defined color connections.\footnote{As we will see, partons with {\tt undefined} color connections are allowed to radiate soft partons into an unrestricted angular region. Since all of our partons are contained in the angular region of the fat jet, this does not cause much of a problem. However, if we wanted to increase the angular region considered in shower deconstruction, we would need to specify color connected partners for all partons.}

\subsection{Kinematics}

We need to describe the kinematics of a splitting of a parton $J$ into two partons, call them $A$ and $B$. There is a big advantage to making the simplest choice for the relation among the corresponding momenta:
\begin{equation}
\label{eq:momentumconservation}
p_J = p_{\LA} + p_{\LB}
\;\;.
\end{equation}
This means that $p_J^2 > 0$ even if $p_A^2 = 0$ and $p_B^2 = 0$. In shower generation (as distinguished from shower deconstruction) one does not do this. One wants $p^2 = 0$ for all intermediate partons since one does not know the virtualities of daughter partons at the time that the splitting is generated. When all partons have $p^2 = 0$, one has to take some momentum from somewhere in order to balance momentum. If we did that for shower deconstruction, the required treatment would be difficult. For shower deconstruction, we simply use Eq.~(\ref{eq:momentumconservation}) and allow all partons to have $p^2 > 0$. Then each parton (or jet) is characterized by four variables, one of which is $\mu^2 \equiv p^2$.

With this choice, each parton is described by four variables: its virtuality $\mu^2$, its rapidity $y$, its azimuthal angle $\phi$, and the absolute value $k$ of its transverse momentum. The $(+,-,1,2)$ components of the momentum of the parton are then\footnote{We use momentum components $p^\pm = (p^0 \pm p^3)/\sqrt 2$.}
\begin{equation}
\label{eq:momentumdecomposition}
p = \left(
\frac{1}{\sqrt 2}\,\sqrt{k^2 + \mu^2}\, e^y,
\frac{1}{\sqrt 2}\,\sqrt{k^2 + \mu^2}\, e^{-y},
k \cos\phi,
k\sin\phi
\right)
\;\;.
\end{equation}

We are now ready to turn to the vertices of our shower history diagrams.

\section{The hard interaction vertex}
\label{sec:hard vertex}

\begin{figure}
\centerline{\includegraphics[width=7.0cm]{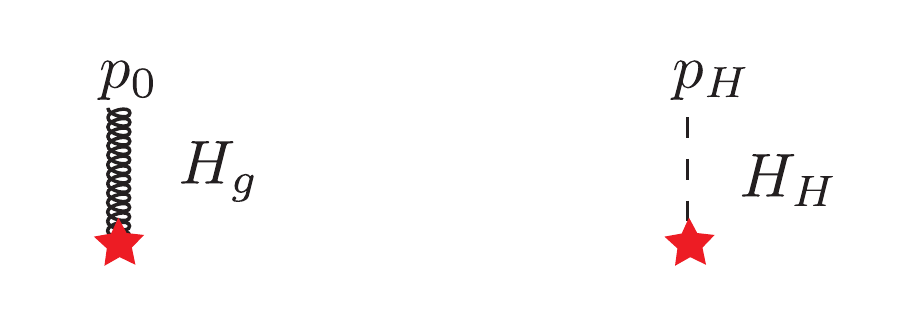}}
\caption{Probability to create the initial parton in the hard interaction. The left hand vertex is for the background process, the right hand vertex is for the signal process.}
\label{fig:HardInteraction}
\end{figure}

We first need a factor to represent the hard scattering process that creates the starting high $p_T$ parton that forms the fat jet, or, more exactly, forms the part of the fat jet that is not from initial state emissions. This factor is represented by the ``star'' vertex, as in Fig.~\ref{fig:HardInteraction}. We consider first the hard vertex for background events.

\subsection{Background}
\label{sec:Hhard_background}

First, we impose a requirement that the scattering process that creates the starting high $p_T$ parton is indeed the dominant hard scattering process in the event. We define $Q^2$ to be the square of the transverse momentum of the fat jet plus the square of its mass,
\begin{equation}
\label{eq:Qsqdef}
Q^2 = 
\left(\sum_{i\in {\rm fat\ jet}} \vec p_{T,i}\right)^{\!\!2}
+\left(\sum_{i\in {\rm fat\ jet}} p_i\right)^{\!\!2}
\;\;.
\end{equation}
We then define $\vec k_{T,\LI}$ to be the transverse momentum of all microjets that are part of the fat jet but are not in the decay products of the initial hard parton. That is, $\vec k_{T,\LI}$ is the transverse momentum of all microjets associated with initial state and underlying event radiation. We demand that
\begin{equation}
\label{eq:IScut}
k_{T,\LI}^2 < Q^2/4
\;\;.
\end{equation}

For the probability density associated with the creation of the initial hard parton, we use a factor
\begin{equation}
\label{eq:Hstart}
H_g = N_{\rm pdf}^g \left(\frac{p_{T,{\rm min}}^2}{k_0^2}\right)^{\!\! N_{\rm pdf}^g}
\frac{1}{k_0^2}\
\Theta(k_{T,\LI}^2 < Q^2/4)
\;\;.
\end{equation}
Here $k_0$ is the transverse momentum of the initial hard parton. The factor $1/k_0^2$ is an approximation to the $k_0^2$ dependence of the square of the hard matrix element. The hard scattering cross section is also proportional to a product of parton distribution functions. We approximate the dependence on the parton distribution functions by including a factor $1/(k_0^2)^{N_{\rm pdf}^g}$, where our default value for the exponent is $N_{\rm pdf}^g = 2$. This value yields an approximation to the one jet inclusive cross section at the Large Hadron Collider, as illustrated in Fig.~11 of ref.~\cite{JetErrors}. The parameter $p_{T,{\rm min}}$ is the smallest allowed transverse momentum of the $Z$-boson against which the initial hard parton recoils, $p_{T,{\rm min}} = 200\ {\rm GeV}$, Eq.~(\ref{eq:pTcut}). The normalization factor $N_{\rm pdf} (p_{T,{\rm min}}^2)^{N_{\rm pdf}}$ is chosen so that the integral $\int dk_0^2\,H$ from $p_{T,{\rm min}}^2$ to infinity is 1. There is an additional normalization factor that we omit because it cancels between the hard scattering cross sections for background and for signal. 

\subsection{Signal}
\label{sec:Hhard_signal}

We also need a factor to represent the hard scattering process that creates the Higgs boson. For this purpose, we use a factor
\begin{equation}
\label{eq:signalstart}
H_H = N_{\rm pdf}^H \left(\frac{p_{T,{\rm min}}^2 + m_H^2}
{k_H^2 + m_H^2}\right)^{\!\! N_{\rm pdf}^H}
\frac{1}{k_H^2 + m_H^2}\
\Theta(k_{T,\LI}^2 < Q^2/4)
\;\;,
\end{equation}
as in Eq.~(\ref{eq:Hstart}). Here $k_H$ is the transverse momentum of the Higgs boson, $m_H$ is the Higgs boson mass, $k_{T,\LI}$ is the total transverse momentum of all partons emitted in the initial state, and $Q^2$ is defined in Eq.~(\ref{eq:Qsqdef}). The remaining factors provide an approximation to the dependence on the parton distribution functions. The default values of the parameters are $N_{\rm pdf}^H = 2$ and $p_{T,{\rm min}} = 200\ {\rm GeV}$ as in Eq.~(\ref{eq:Hstart}). 

\section{Initial state and underlying event radiation}
\label{sec:ISradiation}

\begin{figure}
\centerline{\includegraphics[width=3.0cm]{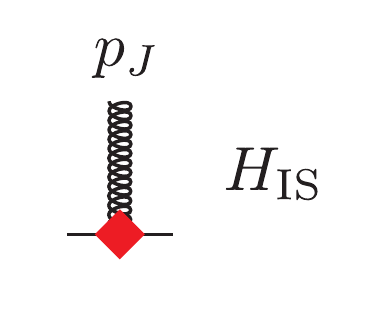}}
\caption{Probability to create a parton by initial state radiation, including both perturbative and nonperturbative radiation.}
\label{fig:ISemission}
\end{figure}

We have seen how to model the hard interaction that creates either a high $p_T$ QCD parton or a Higgs boson. Now we need to model initial state and underlying event radiation, defining an emission probability $H_{\rm IS}$ as illustrated in Fig.~\ref{fig:ISemission}. Consider the probability for the emission of a gluon with positive rapidity from an initial state parton that participates in the hard interaction. Since the gluon has positive rapidity, this emission is predominantly from the active parton ``$\La$'' from hadron A. We use ``$\Lb$'' as the label for the other active incoming quark, from hadron $B$. We take $p_\La$ to be in the $+$ direction and $p_\Lb$ to be in the $-$ direction. We suppose that the emitting parton ``a'' has a color connected partner with label $k$. For the processes that we examine, the initial state partons are likely to be quarks, so there is only one color connected partner. The emitted parton carries the label $J$. As a simple approximation, we assume that it is a gluon. We start with the dipole formula for the squared matrix element for the emission,
\begin{equation}
\label{eq:ISdipolesplitting0}
H_{\rm dipole} \approx \frac{C_\LA}{2}\, (4\pi \as)
\frac{2\, p_\La\cdot p_k}
{p_J \cdot p_\La\ p_J \cdot p_k}
\;\;.
\end{equation}
Writing $p_J \cdot p_k$ in components, this is 
\begin{equation}
\label{eq:ISdipolesplitting1}
H \approx 
\frac{4\pi \as\,C_\LA\, p_\La^+ p_k^-}
{p_\La^+ p_J^- \,(p_J^+ p_k^- + p_J^- p_k^+ - \vec k_{\perp,J}\cdot \vec k_{\perp,k})}
\;\;.
\end{equation}
In order to simplify this, we assume that $p_J^+ p_k^- \gg p_J^- p_k^+$ and $p_J^+ p_k^- \gg |\vec k_{\perp,J}\cdot \vec k_{\perp,k}|$. With this approximation,
\begin{equation}
\label{eq:ISdipolesplitting2}
H \approx 
\frac{8\pi \as\,C_\LA}
{2\,  p_J^- p_J^+ }
\;\;.
\end{equation}
This is exactly
\begin{equation}
\label{eq:ISdipolesplitting3}
H \approx 
\frac{8\pi \as\,C_\LA}{k_J^2 + \mu_J^2}
\;\;.
\end{equation}
This emission probability applies for emitted gluons with positive rapidity, emitted from the active parton in hadron A. It also applies for emitted gluons with negative rapidity, emitted from the active parton in hadron B. To cover all gluons emitted in the central region, we simply use Eq.~(\ref{eq:ISdipolesplitting3}) for both positive and negative rapidity. (We note that $H$ is independent of rapidity with the approximations that we have used.)

In Eq.~(\ref{eq:ISdipolesplitting3}), we choose the squared transverse momentum $k_J^2$ as the argument of $\as$ and we neglect $\mu_J^2$ compared to $k_J^2$:
\begin{equation}
\label{eq:ISdipolesplitting4}
H \approx 
\frac{8\pi \as(k_J^2)\,C_\LA}{k_J^2}
\;\;.
\end{equation}
This expression should then be a fairly good approximation for the emission probability as long as $k_J^2$ is large enough for the emission to be purely perturbative and small enough for the parton momentum fraction carried away by the emitted gluon to be negligible. If the parton momentum fraction carried away by the emitted gluon is not negligible, there should be an additional factor
\begin{equation}
R = \frac{(1-z)\,f(x/(1-z),k_J^2)}{f(x,k_J^2)}
\;\;,
\end{equation}
where $x$ is the momentum fraction of the parton after emitting the gluon, $z x/(1-z)$ is the momentum fraction of the emitted gluon, $x/(1-z)$ is the momentum fraction of the parton before emitting the gluon and the functions $f$ are parton distribution functions. (See Eq.~(8.26) of Ref.~\cite{NSI}). When $k_J^2 \ll Q^2$ we have $z \ll 1$ and $R \approx 1$. However, the approximation $R\approx 1$ breaks down for values of $k_J^2 / Q^2$ at which initial state radiation is still significant. We do not want our simplified shower model to depend on parton distribution functions, so we make a rather crude approximation,
\begin{equation}
R = \frac{1}{(1 + c_R\, k_J/Q)^{n_R}}
\;\;,
\end{equation}
where our default values for the parameters are $c_R = 2$ and $n_R = 1$.

With this factor $R$ included, we should have a fairly good approximation for the emission probability as long as $k_J^2$ is large enough for the emission to be purely perturbative. To give ourselves some flexibility at small $k_J^2$, we replace $k_J^2$ by $k_J^2 + \kappa_\Lp^2$ in the argument of $\as$ and the factor $1/k_J^2$. Our default value for the parameter here is $\kappa_\Lp^2 = 4\ {\rm GeV}^2$. Then the perturbative $H$ is frozen when $k_J$ gets to be much smaller than $\kappa_\Lp$. We then add back a simple non-perturbative function that gives us a chance to adjust the amount of radiation for smaller values of $k_J$.

\begin{figure}
\centerline{\includegraphics[width=8.0cm]{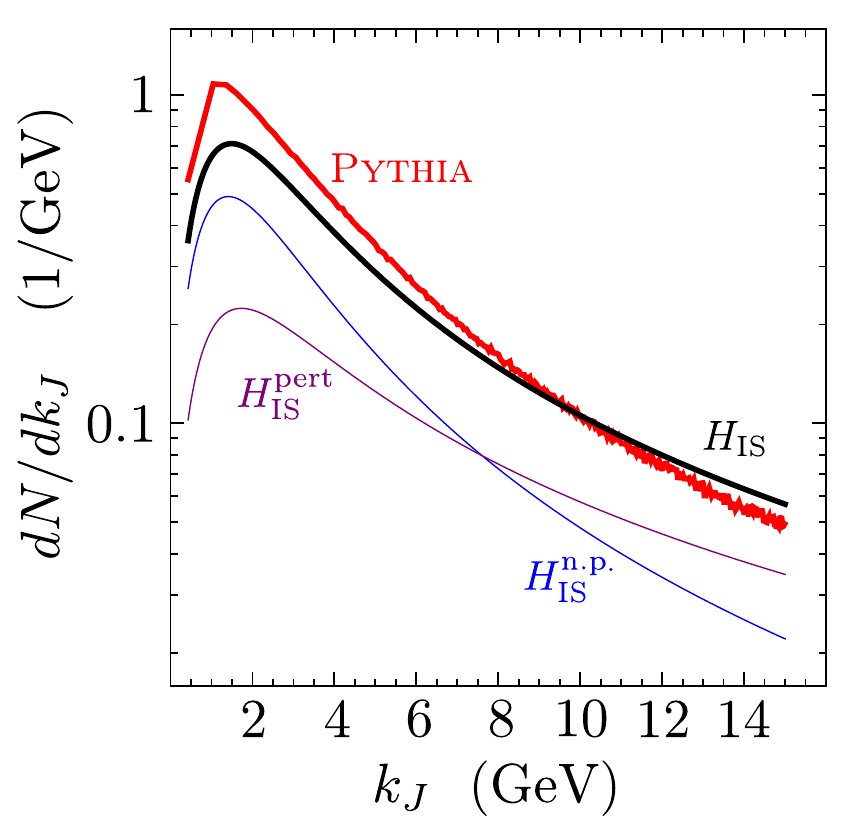}}
\caption{The distribution of initial state jets as a function of their transverse momentum $k_J$ as produced in {\sc Pythia} compared to the distribution produced by $H_{\rm IS}$ and its perturbative and non-perturbative parts. The distributions are integrated over all azimuthal angles and over the rapidity range $-2<y<2$.  For our model, we use $H_{\rm IS}$ from Eq.~(\ref{eq:ISemission}), calling the first term $H_{\rm IS}^{\rm pert}$ and the second term $H_{\rm IS}^{\rm n.p.}$. The distribution from $H_{\rm IS}$ is shown as a heavy line, while the steeper line below is from $H_{\rm IS}^{\rm n.p.}$ while shallower line below is from $H_{\rm IS}^{\rm pert}$.}
\label{fig:UEModelFit}
\end{figure}

This gives the complete initial state emission probability 
\begin{equation}
\label{eq:ISemission}
H_{\rm IS} = 
8\pi\,C_\LA\,
\frac{\as(k_J^2 + \kappa_\Lp^2)}{k_J^2 + \kappa_\Lp^2}\
\frac{1}{(1 + c_R\, k_J/Q)^{n_R}}
+
\frac{16\pi\,c_{\rm np}(\kappa_{\rm np}^2)^{n_{\rm np} - 1}}
{[k_J^2 + \kappa_{\rm np}^2]^{n_{\rm np}}}
\;\;.
\end{equation}
Our default values for the non-perturbative parameters are  $c_{\rm np} = 1$, $\kappa_{\rm np}^2 = 4\ {\rm GeV}^2$, and $n_{\rm np} = 3/2$. It is intended that, with adjustment of parameters, we can include perturbative radiation from the active initial state partons together with radiation at central rapidities and small transverse momenta that is associated with the underlying event and with event pileup. 

Our choice for the parameters is based on comparisons with results from {\sc Pythia}, including the representation in {\sc Pythia} of the effects of the underlying event. We used {\sc Pythia} to produce events for $p + p \to H + Z + X$ where both the Higgs boson and the $Z$-boson decay to muons. For this process, all hadrons are produced by initial state radiation. Although we did not impose a $P_T$ cut on the $Z$-boson, the hard scattering scale here is similar to that for our signal and background processes. We looked for jets that were produced by the initial state radiation, selecting jets using the $k_T$ algorithm with $R = 0.2$ and counting all jets with rapidities in the range $-2 < y < 2$. The resulting distribution as a function of the jet transverse momentum $k_J$ is shown in Fig.~\ref{fig:UEModelFit}. This distribution is to be compared with 
\begin{equation}
\frac{dN_{\rm IS}}{dk_J}=
\int \frac{d^4 p}{(2\pi)^{4}}\ 
2\pi\delta(p^2)\
\delta(|\vec p_T| - k_J)\
\Theta(|y_p| < 2)\
H_{\rm IS}
\;\;.
\end{equation}
This curve, with our choice of parameters, is shown in Fig.~\ref{fig:UEModelFit} along with two more curves corresponding to the two terms in $H_{\rm IS}$. The jets described by $H_{\rm IS}$ are primary jets that can split to produce the jets modeled by {\sc Pythia}, so we have made the primary jet spectrum somewhat harder than the {\sc Pythia} jet spectrum. In Sec.~\ref{sec:results}, we comment on whether the choice of these and other parameters affects the numerical results from shower deconstruction.

\section{Final state QCD shower splittings} 
\label{sec:finalshower}

In this section, we define the main part of the simplified shower, QCD shower splittings.

\subsection{Splitting probability for $g \to g + g$}
\label{sec:gggsplitting}

\begin{figure}
\centerline{\includegraphics[width=4.0cm]{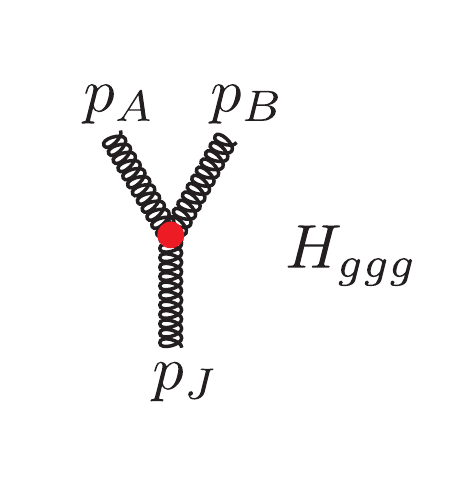}}
\caption{Splitting function for final state $g \to g + g$ splittings.}
\label{fig:Splittingggg}
\end{figure}

The splitting vertex for a QCD splitting $g \to g + g$ is represented by a function $H_{ggg}$ as illustrated in Fig.~\ref{fig:Splittingggg}. We call these the conditional splitting probabilities. Here the condition is that the mother parton has not split already at a higher virtuality.

Let us examine what we should choose for $H_{ggg}$ for a $g \to g + g$ splitting. We take the mother parton to carry the label $J$ and we suppose that the daughter partons are labelled $A$ and $B$ as shown in the figure. The form of the splitting probability depends on which of the two daughter partons is the softer. We let $h$ be the label of the harder daughter parton and $s$ be the label of the softer daughter parton: $k_s < k_h$.

By definition, $k_s < k_h$. We first look at the splitting in the limit $k_s \ll k_h$. The splitting probability is then dominated by graphs in which parton $s$ is emitted from a dipole consisting of parton $J$ and some other parton, call it parton $k$. If $s = A$, then the emitting dipole is formed from parton $h=B$ and parton $k = k(J)_\LL$, while if $s = B$, then the emitting dipole is formed from parton $h=A$ and parton $k = k(J)_\LR$. The choice of $k$ depends on which of the two daughter partons is parton $s$, so where needed we will use the notation $k(s)$ instead of simply $k$.

For $H$, we start with the dipole approximation for the squared matrix element (with $\mu_s^2 = \mu_h^2 = 0$),
\begin{equation}
\label{eq:dipolesplitting}
H_{\rm dipole} \approx \frac{C_\LA}{2}\, (4\pi \as)
\frac{2\, p_h\cdot p_k}
{p_s \cdot p_h\  p_s \cdot p_k}
\;\;.
\end{equation}
We use
\begin{equation}
\begin{split}
2\, p_s \cdot p_h ={}& 2k_s k_h[\cosh(y_s - y_h) - \cos(\phi_s - \phi_h)]
\\ \approx{}& k_s k_h[(y_s - y_h)^2 + (\phi_s - \phi_h)^2]
\\ ={}& k_s k_h\,\theta_{sh}^2
\;\;,
\\
2\, p_s \cdot p_k \approx{}& k_s k_k\,\theta_{sk}^2
\;\;,
\\
2\, p_h \cdot p_k \approx{}& k_h k_k\,\theta_{hk}^2
\;\;,
\end{split}
\end{equation}
where
\begin{equation}
\begin{split}
\label{eq:thetasqdef}
\theta_{sh}^2 ={}& (y_s - y_h)^2 + (\phi_s - \phi_h)^2
\;\;,
\\
\theta_{sk}^2 ={}& (y_s - y_k)^2 + (\phi_s - \phi_k)^2
\;\;,
\\
\theta_{hk}^2 ={}& (y_h - y_k)^2 + (\phi_h - \phi_k)^2
\;\;.
\end{split}
\end{equation}
Thus
\begin{equation}
\label{eq:dipolesplitting2}
H_{\rm dipole} \approx \frac{8\pi\as\,C_\LA}{k_s^2}\ 
\frac{\theta_{hk}^2}
{\theta_{sh}^2\, \theta_{sk}^2}
\;\;.
\end{equation}
This function is singular when parton $s$ is soft, since it is proportional to $1/k_s^2$. It is singular when parton $s$ is parallel to parton $h$. It is also singular when parton $s$ is parallel to parton $k$. We can partition $H_{\rm dipole}$ into two parts, one, $H_{sh}$, associated with emission from parton $h$ and one, $H_{sk}$, associated with emission from parton $k$. (Here we treat parton $s$ as very soft and regard parton $h$ after the emission and parton $J$ before the emission as the same.)

We write
\begin{equation}
\begin{split}
H_{sh} ={}& H_{\rm dipole}\times A'_{hk}
\;\;,
\\
H_{sk} ={}& H_{\rm dipole}\times A'_{kh}
\;\;,
\end{split}
\end{equation}
where
\begin{equation}
\begin{split}
A'_{hk} ={}& \frac{\theta_{sk}^2}
{\theta_{sh}^2 + \theta_{sk}^2}
\;\;,
\\
A'_{kh} ={}& \frac{\theta_{sh}^2}
{\theta_{sh}^2 + \theta_{sk}^2}
\;\;,
\end{split}
\end{equation}
so that
\begin{equation}
A'_{hk} + A'_{kh} = 1
\;\;.
\end{equation}
This dipole partitioning function is that of Ref.~\cite{NSIII}, Eq.~(7.12), adapted to the small angle approximations used here. For a Catani-Seymour dipole shower, one uses a different dipole partitioning function.

With this choice, we have
\begin{equation}
\label{eq:dipolesplitting3}
H_{sh} = \frac{8\pi\as\,C_\LA}{k_s^2}\ 
\frac{\theta_{hk}^2}
{\theta_{sh}^2[\theta_{sh}^2 + \theta_{sk}^2]}
\;\;.
\end{equation}
We can improve this a little so that it works better when parton $s$ is not extremely soft. We recall that, for parton $s$ soft, $\mu_J^2 \approx k_s k_h \theta_{sh}^2$ and that $k_h \approx k_J$ and the angles of parton $J$ are close to those of parton $h$. Thus we take
\begin{equation}
\label{eq:dipolesplitting4}
H_{sh} \approx \frac{8\pi\as\,C_\LA}{\mu_J^2}\,
\frac{k_J^2}{k_s k_h}\,
\frac{\theta_{hk}^2}
{\theta_{sh}^2 + \theta_{sk}^2}
\;\;.
\end{equation}

\begin{figure}
\centerline{\includegraphics[width=10.0cm]{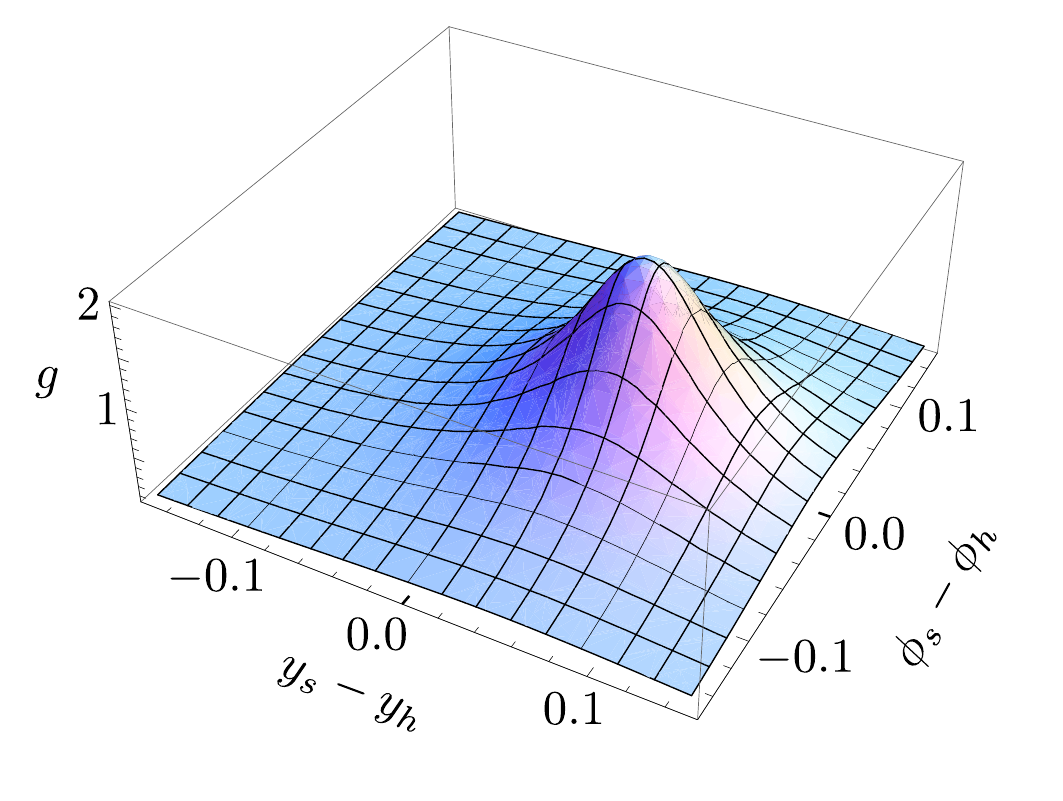}}
\caption{
The angular enhancement factor $g(y_s,\phi_s)$ of Eq.~(\ref{eq:anglefactor}). The coordinates are $(y_s - y_h, \phi_s - \phi_h)$. The color connected parton $k$ is at coordinates $(0.1, 0)$. This figure is adapted from Ref.~\cite{NSII}.
}
\label{fig:angularg}
\end{figure}

The angular factor
\begin{equation}
\label{eq:anglefactor}
g(y_s,\phi_s) = \frac{\theta_{hk}^2}
{\theta_{sh}^2 + \theta_{sk}^2}
\end{equation}
is of some interest. We plot it in Fig.~\ref{fig:angularg}. It enhances radiation into the region between parton $h$ and parton $k$ and disfavors radiation at angles much greater than the angle between parton $h$ and parton $k$. The variable ``pull'' \cite{pull} is designed to separate signal and background events based on this factor. Here, the same effect appears as a natural part of a parton shower based on color dipoles.

So far, we have an approximation that is good in the limit of emission of a soft gluon. This approximation is also good when the gluon labeled $s$ is collinear with the mother parton direction as long as $k_s \ll k_J$. When the two daughter partons are nearly collinear, we have
\begin{equation}
\begin{split}
\frac{k_h}{k_J} \approx{}& z
\;\;,
\\
\frac{k_s}{k_J} \approx{}& 1-z
\;\;,
\end{split}
\end{equation}
where $z$ is the momentum fraction carried by gluon $h$. Our splitting function is proportional to 
\begin{equation}
\frac{k_J^2}{k_s k_h} \approx \frac{1}{z(1-z)}
\;\;.
\end{equation}
This is right for $(1-z) \ll 1$ but it has corrections when $1-z$ is not small. The complete DGLAP splitting kernel for collinear splittings is
\begin{equation}
P_{gg}(z) = 2C_A\ \frac{[1-z(1-z)]^2}{z(1-z)}
\;\;.
\end{equation}
Thus we should replace
\begin{equation}
\frac{k_J^2}{k_s k_h} \to \frac{k_J^2}{k_s k_h}
\left[1 - \frac{k_s k_h}{k_J^2}\right]^2
\;\;.
\end{equation}
Thus we take
\begin{equation}
\label{eq:dipolesplitting5}
H_{sh} \approx \frac{8\pi\as\,C_\LA}{\mu_J^2}\,
\frac{k_J^2}{k_s k_h}\,
\left[1 - \frac{k_s k_h}{k_J^2}\right]^2
\frac{\theta_{hk}^2}
{\theta_{sh}^2 + \theta_{sk}^2}
\;\;.
\end{equation}
%

We need to add another ingredient: $\mu_J^2$ cannot be too large. Suppose that the mother of parton $J$ is parton $K$ and the sister is parton $J'$. We need to be able to neglect $\mu_J^2$ and $\mu_{J'}^2$ in the calculation of $(p_J + p_{J'})^2 \equiv \mu_K^2$. With a little kinematic analysis, we see that neglecting $\mu_J^2$ and $\mu_{J'}^2$ is a good approximation if
\begin{equation}
\begin{split}
\frac{\mu_J^2}{k_J}\, \ll{}& \frac{\mu_K^2}{k_K}
\;\;,
\\
\frac{\mu_{J'}^2}{k_{J'}}\, \ll{}& \frac{\mu_K^2}{k_K}
\;\;.
\end{split}
\end{equation}
We can enforce this condition in an approximate way by requiring
\begin{equation}
\begin{split}
\label{eq:hardnesscut}
2\,\frac{\mu_J^2}{k_J}\, <{}&  \frac{\mu_K^2}{k_K}
\;\;,
\\
2\,\frac{\mu_{J'}^2}{k_{J'}}\, <{}&  \frac{\mu_K^2}{k_K}
\;\;.
\end{split}
\end{equation}
For this reason, we include in $H$ a factor $\Theta(2\mu_J^2/k_J < \mu_K^2/k_K)$. We know $\mu_K^2$ from the shower history. If there is no mother parton because parton $J$ was produced in the hard interaction or by initial state bremsstrahlung, we take $\mu_K^2/k_K = 2 k_J$, so that the virtuality ordering condition becomes simply $\mu_J^2 < k_J^2$.

This same condition, iterated, restricts the daughter virtualities:
\begin{equation}
\begin{split}
2\,\frac{\mu_h^2}{k_h} <{}& \frac{\mu_J^2}{k_J}
\;\;,
\\
2\,\frac{\mu_s^2}{k_s} <{}& \frac{\mu_J^2}{k_J}
\;\;.
\end{split}
\end{equation}

This gives a splitting probability $H$:
\begin{equation}
\label{eq:splittingH}
H_{ggg} = 8\pi C_A\,\frac{\as(\mu_J^2)}{\mu_J^2}\,
\frac{k_J^2}{k_s k_h}
\left[1 - \frac{k_s k_h}{k_J^2}\right]^2
\frac{\theta_{hk}^2}
{\theta_{sh}^2 + \theta_{sk}^2}\
\Theta\!\left(2\,\frac{\mu_J^2}{k_J} < \frac{\mu_K^2}{k_K}\right)
\;\;.
\end{equation}
Here we evaluate $\as$ at the virtuality scale of the splitting. When there is no color connected parton visible, we are forced to simplify this to
\begin{equation}
\label{eq:splittingHnopartner}
H_{\text{no-}k} = 8\pi C_A\,\frac{\as(\mu_J^2)}{\mu_J^2}\,
\frac{k_J^2}{k_s k_h}
\left[1 - \frac{k_s k_h}{k_J^2}\right]^2
\Theta\!\left(2\,\frac{\mu_J^2}{k_J} < \frac{\mu_K^2}{k_K}\right)
\;\;.
\end{equation}
Here there is no restriction on the angles $y_s,\phi_s$ of the emitted soft parton. This is potentially a very bad approximation, but in our case the approximation is tolerable because the emitted soft parton is necessarily within the fat jet. When, in addition, there is no mother parton $K$, this becomes
\begin{equation}
\label{eq:splittingHnomother}
H_{\text{no-}K} = 8\pi C_A\,\frac{\as(\mu_J^2)}{\mu_J^2}\,
\frac{k_J^2}{k_s k_h}
\left[1 - \frac{k_s k_h}{k_J^2}\right]^2
\Theta\!\left(\mu_J^2 < k_J^2\right)
\;\;.
\end{equation}

\subsection{Splitting probability for $q \to q + g$ and $\bar q \to \bar q + g$}
\label{sec:bbgsplitting}

\begin{figure}
\centerline{\includegraphics[width=7.0cm]{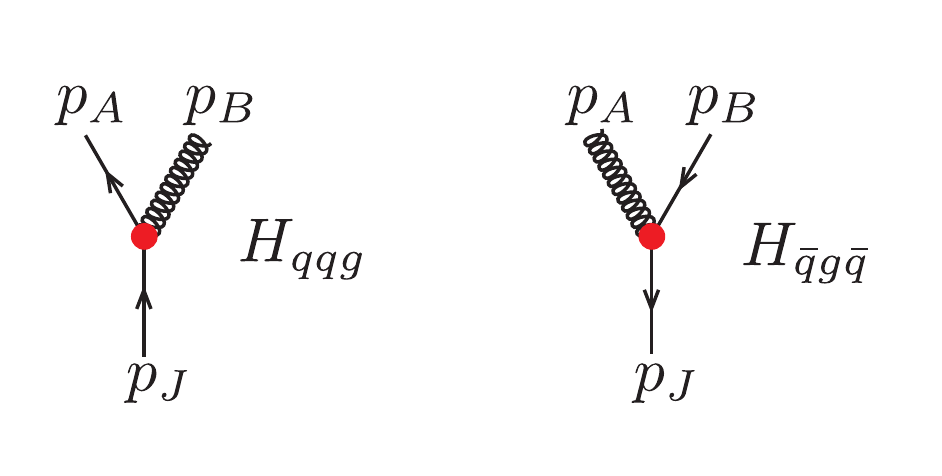}}
\caption{Splitting functions for final state QCD splittings of a quark or antiquark, including a $b$ or $\bar b$ quark.}
\label{fig:Splittingqgq}
\end{figure}

Quarks and antiquarks can radiate gluons. These splittings are represented by the splitting probabilities $H_{qqg}$ and $H_{\bar q g \bar q}$ that are illustrated in Fig.~\ref{fig:Splittingqgq}. We treat the splitting of a bottom quark as identical to the splitting of a light quark, neglecting the bottom quark mass. We take the splitting probability to be
\begin{equation}
\label{eq:splittingHqqg}
H_{qqg} = H_{\bar q g \bar q} =
8\pi C_\LF\,
\frac{\as(\mu_J^2)}{\mu_J^2}\  \frac{k_J}{k_g}
\left[
1 + \left(\frac{k_q}{k_J}\right)^2
\right]
\frac{\theta_{qk}^2}
{\theta_{gq}^2 + \theta_{gk}^2}\,
\Theta\!\left(2\,\frac{\mu_J^2}{k_J} < \frac{\mu^2_K}{k_K}\right)
\;\;.
\end{equation}
The derivation follows the derivation that led to Eq.~(\ref{eq:splittingH}). Here $k_g$ is the transverse momentum of the gluon, $k_q$ is the transverse momentum of the quark or antiquark, and $k_J$ is the transverse momentum of the mother quark. Then using $k_q/k_J \approx z$ and $k_g/k_J \approx (1-z)$, the factor containing these ratios gives the collinear splitting function
\begin{equation}
\label{eq:Pqq}
P_{qq} = C_F\, \frac{1 + z^2}{1-z}
\end{equation}
in the collinear limit. 

There is an angle factor in which $q$ labels daughter quark or antiquark, $g$ labels the emitted gluon, and $k$ labels the color connected partner of the quark or antiquark. If there is no color connected partner in the fat jet, this angle factor is to be omitted.

There is a theta function that restricts the mass $\mu_J^2$ of the daughter pair to be less than $\mu_K^2 k_J/(2 k_K)$, where $K$ labels the mother of parton $J$. With our approximations for shower histories, a quark or antiquark always has a mother parton.

\subsection{Splitting probability for $g \to q + \bar q$}
\label{sec:gqqsplitting}

\begin{figure}
\centerline{\includegraphics[width=4.0cm]{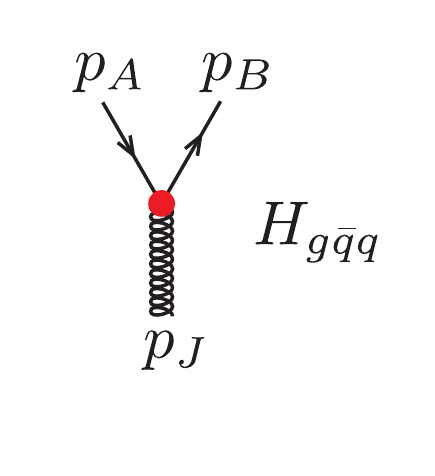}}
\caption{Splitting function for final state QCD splittings that produce a $q\bar q$ pair.}
\label{fig:Splittinggqq}
\end{figure}

We need one more QCD splitting probability, for $g \to q + \bar q$, including $g \to b + \bar b$ as illustrated in Fig.~\ref{fig:Splittinggqq}.  Note that this splitting is important because $g \to b + \bar b$ is the main background for the $H \to b + \bar b$ signal, so we need to keep track of $g \to b + \bar b$ splittings even if they have a small probability.

To construct the splitting function that we need, we can start with the $q \to q + g$ splitting function in Eq.~(\ref{eq:splittingHqqg}). We can take the collinear limit, setting the angle factor to 1. Then we replace $P_{qq}$, Eq.~(\ref{eq:Pqq}), with $z \approx k_q/k_J$ and $(1-z) \approx k_{g}/k_J$, by
\begin{equation}
P_{qg} = T_\LR\, [z^2 + (1-z)^2]
\end{equation}
with $z \approx k_q/k_J$ and $(1-z) \approx k_{\bar q}/k_J$. This gives
\begin{equation}
\label{eq:splittinggtob}
H_{g\bar q q} = 8\pi T_\LR\,
\frac{\as(\mu_J^2)}{\mu_J^2}
\frac{k_q^2 + k_{\bar q}^2}{k_J^2}\
\Theta\!\left(2\,\frac{\mu_J^2}{k_J} < \frac{\mu_K^2}{k_K}\right)
\;\;.
\end{equation}
Note that this function is big for small $\mu_J^2$ in the limit in which the quark pair is collinear, but that there is no additional singularity when the quark or antiquark is soft.

For a gluon splitting to $b + \bar b$ we use $H_{g\bar b b} = H_{g\bar q q}$ as given above. For a gluon splitting to $(u,\bar u)$, $(d,\bar d)$, $(s,\bar s)$, and $(c,\bar c)$, we include all four cases at once by using $(n_\Lf - 1) H_{g\bar q q}$, where $(n_\Lf - 1) = 4$.

There is a theta function that restricts the mass $\mu_J^2$ of the daughter pair to be less than $\mu_K^2 k_J/(2 k_K)$, where $K$ labels the mother of parton $J$. If there is no mother parton $K$, this theta function becomes $\Theta\!\left(\mu_J^2 < k_J^2\right)$.

\section{The Sudakov factor in the final state shower}
\label{sec:Sudakov}

We have given definitions for splitting probabilities in the simplified shower. An important part of a parton shower event generator is the probability that a parton that was created at a virtuality scale $\mu_K^2$ has not split before it finally does split at a scale $\mu_J^2$. This is the Sudakov factor and has the form $\exp(-S)$, where $S$ is the integral of the splitting probability down to the scale $\mu_J^2$. In this section, we explore how to approximate $S$.

\subsection{Variables for parton splitting}
\label{sec:splittingvariables}

To evaluate the Sudakov exponent, we need to understand in some detail the integrations for combining two partons. 

We use
\begin{equation}
\label{eq:d4ptomukyphi}
\int\!\frac{d^4 p}{(2\pi)^4}\ \cdots
= \int_0^\infty\!\frac{d\mu^2}{2\pi}\
\frac{1}{4(2\pi)^3}
\int_0^\infty\!dk^2
\int_{-\infty}^{\infty}\!dy
\int_0^{2\pi}\!d\phi\
\cdots
\;\;.
\end{equation}
We consider integrations over the momenta of partons A and B that we would like to combine to make parton $J$:
\begin{equation}
\label{eq:Idef}
I = 
\int\!\frac{d\mu_A^2}{2\pi}\
\frac{1}{4(2\pi)^3}
\int\!dk_A^2
\int\!dy_A
\int\!d\phi_A\
\int\!\frac{d\mu_B^2}{2\pi}\
\frac{1}{4(2\pi)^3}
\int\!dk_B^2
\int\!dy_B
\int\!d\phi_B\
\cdots
\;\;.
\end{equation}
Now we insert
\begin{equation}
1 = \int \!\frac{d^4 p_J}{(2\pi)^4}\ (2\pi)^4\delta^4(p_A + p_B - p_J)
\end{equation}
and use Eq.~(\ref{eq:d4ptomukyphi}) for $\int d^4 p_J$. This gives
\begin{equation}
\begin{split}
I ={}& \frac{1}{4(2\pi)^3}
\int\!dk_J^2
\int\!dy_J
\int\!d\phi_J
\int\!\frac{d\mu_J^2}{2\pi}\
\int\!\frac{d\mu_A^2}{2\pi}\int\!\frac{d\mu_B^2}{2\pi}
\\ &\times
\frac{1}{(2\pi)^2}\,\frac{1}{16}
\int\!dk_A^2
\int\!dy_A
\int\!d\phi_A
\int\!dk_B^2
\int\!dy_B
\int\!d\phi_B
\\ & \times
\delta^4(p_A + p_B - p_J)
\cdots
\;\;.
\end{split}
\end{equation}
In the second line, we have six variables, $k_A$, $y_A$, $\phi_A$, $k_B$, $y_B$, and $\phi_B$, restricted by four delta functions. This leaves an integration over two variables. We choose one of the variables to be the momentum fraction
\begin{equation}
\begin{split}
\label{eq:zdef0}
z ={}& \frac{k_A}{k_A + k_B}
\;\;,
\\
1-z ={}& \frac{k_B}{k_A + k_B}
\;\;.
\end{split}
\end{equation}
For the other integration variable describing the splitting, we use $\varphi$ defined by
\begin{equation}
\label{eq:tanvarphi0}
\tan\varphi =
\frac{\sinh(\Delta y/2)\cos(\Delta\phi/2)}{\cosh(\Delta y/2)\sin(\Delta\phi/2)}
\;\;,
\end{equation}
where
\begin{equation}
\begin{split}
\Delta y ={}& y_A - y_B
\;\;,
\\
\Delta \phi ={}& \phi_A - \phi_B
\;\;.
\end{split}
\end{equation}
Thus $\varphi$ is approximately the angle about the origin in the $(\Delta \phi,\Delta y)$ plane.

Then
\begin{equation}
\begin{split}
\label{eq:Iresult}
I ={}& \int\!\frac{d\mu_A^2}{2\pi}\int\!\frac{d\mu_B^2}{2\pi}\
\frac{1}{4(2\pi)^3}\int\!dk_J^2
\int\!dy_J
\int\!d\phi_J
\\&\times
\frac{1}{4(2\pi)^2}
\int\!\frac{d\mu_J^2}{2\pi}
\int\!dz \int\! d\varphi\
J
\cdots
\;\;,
\end{split}
\end{equation}
where $J$ is a jacobian to be discussed presently. We think about this as follows. We combine two subjets $A$ and $B$ of mass $\mu_A$ and $\mu_B$. We display integrations over $\mu_A$ and $\mu_B$, but these integrations remain unaltered between the original integral (\ref{eq:Idef}) and the result Eq.~(\ref{eq:Iresult}). In the original integral, we integrate over $k^2$, $y$ and $\phi$ for the two constituent jets, with the standard factor\footnote{The Feynman rules that we use for calculating squared matrix elements assume that momentum integrations are $(2\pi)^{-4}\int\! d^4p\  (2\pi)\,\delta(p^2 - \mu^2)$, which gives this factor to accompany integrations over $k^2$, $y$, and $\phi$ as in Eq.~(\ref{eq:d4ptomukyphi}).} $1/[4 (2\pi)^3]$ for each. The subjets are combined to make a jet $J$ described by $k_J^2$, $y_J$ and $\phi_J$. We integrate over these variables with the standard factor $1/[4 (2\pi)^3]$. This leaves variables $\mu_J$, $z$ and $\varphi$ that describe the splitting. Integration over these variables comes with a factor $1/[4(2\pi)^3]$ and a jacobian $J$.

In Eq.~(\ref{eq:Iresult}), a ``strong ordering'' approximation applies for jet masses, $\mu_A \ll \mu_J$ and $\mu_B \ll \mu_J$. In turn, $\mu_J$ is small compared to $k_A$, $k_B$ and $k_J$. For this reason, it is a sufficient approximation to set $\mu_A = \mu_B = 0$ in $J$. In the appendix of this paper, we calculate $J$ with $\mu_A = \mu_B = 0$. We find a quite simple result, 
\begin{equation}
J = \frac{\sinh^2(\Delta y/2) + (1 + \mu_J^2/k_J^2)\sin^2(\Delta\phi/2)}
{\sinh^2(\Delta y/2)\cosh^2(\Delta y/2) 
+ (1 + \mu_J^2/k_J^2)\sin^2(\Delta\phi/2)\cos^2(\Delta\phi/2)}
\;\;.
\end{equation}
This result is even simpler when $\Delta y$ and $\Delta\phi$ are small. Since $\cosh(\Delta y/2) \approx 1$ and  $\cos(\Delta\phi/2) \approx 1$ for small angles, we have
\begin{equation}
J \approx 1
\end{equation}
for small angles.

\subsection{Splitting probability and the Sudakov exponent}

We will insert a splitting probability into each integration over the splitting variables, so that the splitting probability differential in the splitting variables $\mu_J^2, z ,\varphi$ is
\begin{equation}
\begin{split}
\label{eq:DifferentialSplittingProbability}
d{\cal P} = {}&
\frac{1}{4(2\pi)^{3}}
d\mu_J^2\ dz\ d\varphi\
H e^{-S}
\end{split}
\end{equation}
Here we have approximated the jacobian $J$ by its small angle form, $J\approx 1$. We also use small angle approximations in $H$, as in our expressions in Sec.~\ref{sec:finalshower}. For instance, we take $k_A/k_J \approx z$ and $k_B/k_J \approx (1-z)$.

The corresponding total splitting probability is
\begin{equation}
\begin{split}
\label{eq:splitprobability}
\int d{\cal P} = {}&
\frac{1}{4(2\pi)^3}
\int\!d\mu_J^2
\int\!dz \int\!d\varphi\
H e^{-S}
\;\;.
\end{split}
\end{equation}
Here $H$ is the conditional splitting probability for a mother parton to split if it has not split at a higher virtuality than $\mu_J^2$ and $e^{-S}$ is the probability, derived from $H$, that the mother parton has not split at a higher virtuality. Given the physical meaning of the Sudakov factor, one would like
\begin{equation}
\label{eq:SudakovExponent0}
S \approx \frac{1}{4(2\pi)^3}
\int\!d\bar\mu_J^2 \,\Theta(\mu_J^2 < \bar\mu_J^2)
\int\!d\bar z \int\!d\bar\varphi\
H(\bar p_A,\bar p_B)\,
\Theta(\{\bar p_A,\bar p_B\} \in {\rm fat\ jet})
\;\;.
\end{equation}
Here $\bar p_A$ and $\bar p_B$ denote the momenta of the daughter partons in a possible splitting and $\bar\mu_J^2$, $\Delta \bar y$, and $\Delta \bar\phi$ denote parameters of the possible splitting.

The theta function $\Theta(\{\bar p_A,\bar p_B\} \in {\rm fat\ jet})$ is present for the following reason. Parton $J$ has, in each interval of virtuality $d\bar\mu_J^2$, a probability to emit a soft, wide angle gluon that is not seen because it is outside the boundary of the fat jet. The probability for emission of such a ghost gluon is most substantial when the color connected partner for the emission is itself outside the fat jet. Fortunately, the momentum of the emitted ghost gluon is small, since it must be a soft, wide angle gluon. Thus it is a sensible approximation to ignore this momentum loss. Since we cannot see the ghost emissions, we ignore them completely. This means that we ignore them in the Sudakov exponent $S$ by integrating only over splittings in which both daughter partons are in the fat jet.

\subsection{Sudakov exponent for gluon splitting}
\label{sec:sudakov}

\begin{figure}
\centerline{\includegraphics[width=4.0cm]{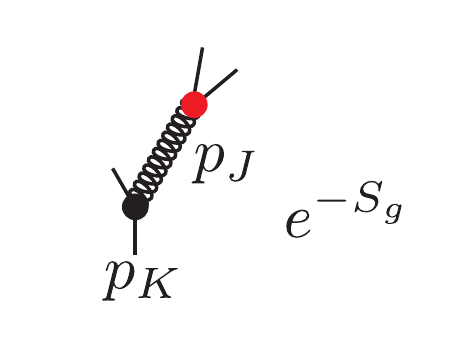}}
\caption{Sudakov factor between final state splittings for a gluon.}
\label{fig:Sudakovggg}
\end{figure}

As stated in the previous subsection, the Sudakov factor is the probability that the mother parton $J$ did not split at a virtuality above $\mu_J^2$. Thus the Sudakov factor is $\exp(-S)$, where $S$ is the probability for the mother parton to have split at a value of $\mu_J$ that is greater than the value at which the splitting did, in fact, occur. The corresponding Sudakov factors are associated with the propagators in our shower history diagrams. For instance, for a gluon, the factor $\exp(-S_g)$ is indicated in Fig.~\ref{fig:Sudakovggg}. There are three contributions to $S_g$, corresponding to $g \to g + g$, $g \to q + \bar q$, and $g \to b + \bar b$. Note that the total $S_g$ appears in $\exp(-S_g)$ independently of whether the gluon ultimately decays to $g + g$, $q + \bar q$, or $b + \bar b$. In this section, we work out the contribution from $g \to g + g$.

We start with in Eq.~(\ref{eq:splittingH}) for $H_{ggg}$ with $k_s/k_J$ replaced by $z$ and $k_h/k_J$ replaced by $(1-z)$ in the case that the label $s$ of the softer daughter parton is $s=A$ or $k_h/k_J$ replaced by $z$ and $k_s/k_J$ replaced by $(1-z)$ in the case that $s=B$. Since $H$ is symmetric under $k_s \leftrightarrow k_h$, the choice of $s$ does not affect the form of the result. However, now $s = A$ corresponds to $z < 1/2$ and $s = B$ corresponds to $z > 1/2$. This gives
\begin{equation}
\label{eq:H0}
H_{ggg} \approx 8\pi C_\LA\,\frac{\as(\bar\mu_J^2)}{\bar\mu_J^2}\,
\frac{[1 - z(1-z)]^2}{z (1-z)}
\frac{\theta_{hk}^2}
{\theta_{sh}^2 + \theta_{sk}^2}\
\Theta\!\left(2\,\frac{\bar\mu_J^2}{k_J} < \frac{\mu_K^2}{k_K}\right)
\;\;.
\end{equation}

In the angular factor ${\theta_{hk}^2}/ {[\theta_{sh}^2 + \theta_{sk}^2]}$, we use the notation from Eq.~(\ref{eq:thetasqdef}) that $\theta_{\alpha\beta}^2 = (y_\alpha - y_\beta)^2 + (\phi_\alpha - \phi_\beta)^2$. The angular factor is one for small angles $\theta_{sh}$ and is small when $\theta_{sh} \gg \theta_{hk}^2$. Thus it is approximately a theta function that requires $\theta_{sh} < \theta_{hk}$. Here $\theta_{hk}$ is approximately the angle $\theta_{k(s)}$ between the mother parton and the parton $k(s)$ that carries the color line of the mother parton that is carried by the emitted soft parton. Thus we replace
\begin{equation}
\label{eq:anglereplacement}
\frac{\theta_{hk}^2}
{\theta_{sh}^2 + \theta_{sk}^2}
\to \Theta(\theta < \theta_{k(s)})
\;\;.
\end{equation}
This is the angle-ordering approximation to the true dipole matrix element \cite{MarchesiniWebber}. It is a rather crude approximation locally in angle space, but is a pretty good approximation after integrating from large $\theta$ to small $\theta$. With this approximation, we have
\begin{equation}
\label{eq:H1}
H_{ggg} \approx 8\pi C_\LA\,\frac{\as(\bar\mu_J^2)}{\bar\mu_J^2}\,
\frac{[1 - z(1-z)]^2}{z (1-z)}
\Theta\big(\theta_{sh}^2 < \theta_{k(\bar s)}^2\big)\,
\Theta\!\left(2\,\frac{\bar \mu_J^2}{k_J} < \frac{\mu_K^2}{k_K}\right)
\;\;.
\end{equation}
We can translate the restrictions on $\theta_{sh}$ to restrictions on $z$. From Eq.~(\ref{eq:z1mz}) of the appendix, we have, in the limit of small angles,
\begin{equation}
\label{eq:mutotheta}
\frac{\mu_J^2}{k_J^2} \approx z(1-z)\, \theta^2_{sh}
\;\;.
\end{equation}
Thus for $z < 1/2$ the relation $\theta_{sh}^2 < \theta_{k(\bar s)}^2$ becomes
\begin{equation}
z(1-z) > \frac{1}{\theta_{k(A)}^2}\,
\frac{\bar \mu_J^2}{k_J^2}
\;\;.
\end{equation}
Presuming that the right hand side of this inequality is much smaller than 1, we can simplify this approximately to
\begin{equation}
z > \frac{1}{\theta_{k(A)}^2}\,
\frac{\bar \mu_J^2}{k_J^2}
\;\;.
\end{equation}
Similarly, we have a restriction on how small $(1-z)$ can be,
\begin{equation}
(1-z) > \frac{1}{\theta_{k(B)}^2}\,
\frac{\bar \mu_J^2}{k_J^2}
\;\;.
\end{equation}
These inequalities can be combined as
\begin{equation}
\frac{1}{\theta_{k(A)}^2}\,
\frac{\bar \mu_J^2}{k_J^2}
< z 
< 1 - \frac{1}{\theta_{k(B)}^2}\,
\frac{\bar \mu_J^2}{k_J^2}
\;\;.
\end{equation}
Thus
\begin{equation}
\begin{split}
\label{eq:H2}
H_{ggg} \approx {}& 8\pi C_\LA\,\frac{ \as(\bar\mu_J^2)}{\bar\mu_J^2}\,
\frac{[1 - z(1-z)]^2}{z (1-z)}
\\ & \times
\Theta\!\left(\frac{1}{\theta_{k(A)}^2}\,
\frac{\bar \mu_J^2}{k_J^2}
< z 
< 1 - \frac{1}{\theta_{k(B)}^2}\,
\frac{\bar \mu_J^2}{k_J^2}\right)\,
\Theta\!\left(2\,\frac{\bar\mu_J^2}{k_J} < \frac{\mu_K^2}{k_K}\right)
\;\;.
\end{split}
\end{equation}

For the theta function $\Theta(\{\bar p_A,\bar p_B\} \in {\rm fat\ jet})$ in Eq.~(\ref{eq:SudakovExponent0}), we note that if $\theta_{k(s)}$ is much smaller than the fat jet radius $R_\LF$, the theta function that imposes angular ordering, $\Theta(\theta^2 < \theta_{k(s)}^2)$, will almost always enforce that $\bar p_A$ and $\bar p_B$ are in the fat jet, so that $\Theta(\{\bar p_A,\bar p_B\} \in {\rm fat\ jet}) = 1$. On the other hand, sometimes there is no color connected parton with label $k(\bar s)$ in the fat jet. Then we use Eq.~(\ref{eq:splittingHnopartner}), which effectively defines $\theta_{k(\bar s)} = \infty$. In this case, the theta function $\Theta(\{\bar p_A,\bar p_B\} \in {\rm fat\ jet})$ limits $\theta$ to a maximum value on the order of the fat jet radius $R_\LF$. We take a simple approximation and replace $\Theta(\{\bar p_A,\bar p_B\} \in {\rm fat\ jet})$ by $\Theta(\theta^2 < R_0^2)$, where $R_0$ is an adjustable parameter with default value $R_0 = R_\LF$. Thus we understand that we should make the replacement
\begin{equation}
\label{eq:nokreplacement}
\theta_{k(\bar s)} \to R_0
\end{equation}
when there is no color connected parton $k(s)$.

In the case that parton $J$ is the parton that has no mother parton $K$ because it originates a jet, we use Eq.~(\ref{eq:splittingHnomother}) for $H$. This amounts to making the replacement
\begin{equation}
\label{eq:noKreplacement}
\frac{\mu_K^2}{2k_K} \to k_J
\end{equation}
when there is no mother parton $K$.

With these approximations for $H$, can insert $H_{ggg}$ into Eq.~(\ref{eq:SudakovExponent0}) to obtain
\begin{equation}
\begin{split}
\label{eq:SudakovExponent4}
S_{ggg} \approx{}&
\int\! \frac{d\bar\mu_J^2}{\bar\mu_J^2}\
\Theta\!\left(\mu_J^2 < \bar\mu_J^2 < \frac{k_J}{2k_K}\,\mu_K^2\right)\,
\frac{\as(\bar\mu_J^2)}{2\pi}
\\&\times
\int\!dz\
\Theta\!\left(\frac{1}{\theta_{k(A)}^2}\,
\frac{\bar \mu_J^2}{k_J^2}
< z 
< 1 - \frac{1}{\theta_{k(B)}^2}\,
\frac{\bar \mu_J^2}{k_J^2}\right)\,
C_\LA\,
\frac{[1 - z(1-z)]^2}{z (1-z)}
\;\;,
\end{split}
\end{equation}
where we understand that we are to make the replacement (\ref{eq:nokreplacement}) in the case that there is no color connected parton $k(s)$ and the replacement (\ref{eq:noKreplacement}) in the case that there is no mother parton $K$. Here we have performed the integration over $\varphi$ since, with our approximations, the integrand does not depend on $\varphi$.

Note the structure of this. We integrate half the DGLAP kernel over $\bar\mu_J^2$ and $z$, with limits on the $z$ integral from the angular ordering approximation to the quantum coherence of soft gluon emission from color dipoles. We have half of the DGLAP kernel for $g \to g + g$ because we are integrating over the phase space for two identical particles and need a statistical factor 1/2.

We can perform the integration over $z$, giving
\begin{equation}
\begin{split}
\label{eq:SudakovExponent5}
S_{ggg} \approx{}&
2 C_\LA
\int\! \frac{d\bar\mu_J^2}{\bar\mu_J^2}\
\Theta\!\left(\mu_J^2 < \bar\mu_J^2 < \frac{k_J}{2k_K}\,\mu_K^2\right)\,
\frac{\as(\bar\mu_J^2)}{2\pi}
\left[
\log\left(\theta_{k(A)}\theta_{k(B)}k_J^2/\bar \mu_J^2\right)
- \frac{11}{12}
\right]
\;\;.
\end{split}
\end{equation}
Here we have omitted terms that are suppressed by a power of $\bar \mu_J^2/[k_J^2 \theta_{k(A)}^2]$ or $\bar \mu_J^2/[k_J^2 \theta_{k(B)}^2]$.

We can perform the integration over $\bar \mu$ by changing variables to $\as$ using
\begin{equation}
\label{eq:asevolution}
\frac{d\as(\mu^2)}{d\log(\mu^2)}
= - b_0\, \as(\mu^2)^2
\;\;,
\end{equation}
where $b_0 = ({33 - 2 n_\Lf})/({12\pi})$. We take the number of flavors to be $n_\Lf = 5$. We write
\begin{equation}
\label{eq:asresult}
\log\left(
\frac{\theta_{k(A)}\theta_{k(B)}k_J^2}{\bar\mu_J^2}\right)
= \frac{1}{b_0}
\left[
\frac{1}{\as\!\left(\theta_{k(A)}\theta_{k(B)} k_J^2 \right)}
-
\frac{1}{\as\!\left(\bar\mu_J^2\right)}
\right]
\;\;.
\end{equation}
This gives
\begin{equation}
\begin{split}
\label{eq:SudakovExponent6}
S_{ggg} \approx{}&
\frac{C_\LA}{\pi b_0^2}
\biggl\{
\log\!\left(\frac{\as\!\left(\mu_J^2\right)}
{\as\!\left(k_J\mu_K^2/(2 k_K)\right)}\right)
\left[\frac{1}{\as(\theta_{k(A)}\theta_{k(B)}k_J^2)} 
- \frac{11 b_0}{12}
\right]
\\&\quad
+ \frac{1}{\as\!\left(\mu_J^2\right)}
- \frac{1}{\as\!\left( k_J\mu_K^2/(2 k_K)\right)}
\biggr\}
\;\;.
\end{split}
\end{equation}
Since $\mu_J^2 < k_J\mu_K^2/(2 k_K)$, this quantity is positive as long as the partner angles $\theta_{k(A)}$ and $\theta_{k(A)}$ are not too small. However, since $S < 0$ is unphysical, we replace $S_{ggg} \to S_{ggg} \Theta(S_{ggg} > 0)$ just to be sure that we are never enhancing an unphysical region by having $e^{-S} > 1$.

We also evaluate the Sudakov exponent for a $g \to q + \bar q$ splitting. Here we use $H_{g\bar q q}$ from Eq.~(\ref{eq:splittinggtob}). This gives
\begin{equation}
\label{eq:SudakovExponentgqq1}
S_{g\bar q q} \approx \int\! \frac{d\bar\mu_J^2}{\bar\mu_J^2}\,
\Theta\!\left(\mu_J^2 < \bar\mu_J^2 < \frac{k_J}{2k_K}\,\mu_K^2\right)
\frac{\as(\mu_J^2) }{2\pi}
\int\!dz\
\,T_\LR \,
[{z^2 + (1-z)^2}]
\;\;.
\end{equation}
We can perform the $z$-integration to give
\begin{equation}
\label{eq:SudakovExponentgqq2}
S_{g\bar q q} \approx  \frac{2T_\LR}{3} \int\! d\bar\mu_J^2\,
\Theta\!\left(\mu_J^2 < \bar\mu_J^2 < \frac{k_J}{2k_K}\,\mu_K^2\right)
\frac{\as(\mu_J^2) }{2\pi}\,
\frac{1}{\bar\mu_J^2}
\;\;.
\end{equation}
Then, we can perform the $\bar \mu^2$ integration using Eq.~(\ref{eq:asevolution}) to give
\begin{equation}
\label{eq:SudakovExponentgqq3}
S_{g\bar q q} \approx \frac{T_\LR}{3 \pi b_0}\,
\log\!\left(\frac{\as(\mu_J^2)}{\as(k_J\mu_K^2/(2 k_K))}\right)
\;\;.
\end{equation}

Adding $S_{ggg}$ and one copy of $S_{g\bar q q}$ for each quark flavor, including the $b$-quark, we obtain the complete Sudakov exponent for gluon splitting
\begin{equation}
\label{eq:totalSg}
S_g = S_{ggg}\, \Theta(S_{ggg} > 0)
+ n_\Lf S_{g\bar q q}
\;\;.
\end{equation}
%

\begin{figure}
\centerline{\includegraphics[width=7.0cm]{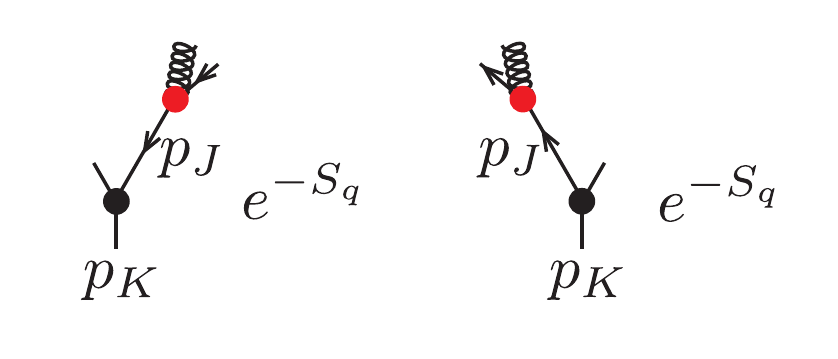}}
\caption{Sudakov factor between final state emission of a gluon from a quark or antiquark. The quark or antiquark flavor can be $b$ or $u$, $d$, $s$ or $c$. The previous splitting can be either a gluon emission, a $g \to q + \bar q$ or $g \to b + \bar b$ splitting or a Higgs boson decay to $b + \bar b$.}
\label{fig:Sudakovq}
\end{figure}

\subsection{Sudakov exponent for quark splitting}
\label{sec:sudakovb}

The Sudakov factor for a quark splitting is illustrated in Fig.~\ref{fig:Sudakovq}. The corresponding Sudakov exponent is given by Eq.~(\ref{eq:SudakovExponent0}) using $H_{qqg}$ from Eq.~(\ref{eq:splittingHqqg}). In $H_{qqg}$ we replace the angular factor ${\theta_{qk}^2}/
[{\theta_{gq}^2 + \theta_{gk}^2}]$ by $\Theta(\theta < \theta_{k})$ as in Eq.~(\ref{eq:splittingHqqg}). In turn, the restriction on $\theta$ amounts to a restriction on $z$,
\begin{equation}
(1-z) > \frac{1}{\theta_{k}^2}\,
\frac{\bar \mu_J^2}{k_J^2}
\;\;.
\end{equation}
This gives
\begin{equation}
\begin{split}
\label{eq:SudakovExponentqqg1}
S_{qqg} \approx {}& 
\int\! \frac{d\bar\mu_J^2}{\bar\mu_J^2}\,
\Theta\!\left(\mu_J^2 < \bar\mu_J^2 < \frac{k_J}{2k_K}\,\mu_K^2\right)
\frac{\as(\bar\mu_J^2)}{2\pi}
\int\!dz\
\Theta\left((1-z) > \frac{1}{\theta_{k}^2}\,
\frac{\bar \mu_J^2}{k_J^2}\right)
\\&\times
C_\LF\,
\frac{1+z^2}{1-z}
\;\;.
\end{split}
\end{equation}
We can perform the $z$-integration to obtain
\begin{equation}
\begin{split}
\label{eq:SudakovExponentqqg2}
S_{qqg} \approx {}& 
2C_\LF \int\! \frac{d\bar\mu_J^2}{\bar\mu_J^2}\,
\Theta\!\left(\mu_J^2 < \bar\mu_J^2 < \frac{k_J}{2k_K}\,\mu_K^2\right)
\frac{\as(\bar\mu_J^2)}{2\pi}\ 
\left[\log\left(\theta_k^2\,{k_J^2}/{\bar \mu_J^2}\right) - \frac{3}{4}
\right]
\;\;.
\end{split}
\end{equation}
Here we have neglected terms suppressed by a power of ${\bar \mu_J^2}/({k_J^2}\theta_k^2)$.

We can now use Eqs.~(\ref{eq:asevolution}) and (\ref{eq:asresult}) to perform the $\bar \mu_J^2$ integration, giving
\begin{equation}
\begin{split}
\label{eq:SudakovExponentqqg3}
S_{qqg} \approx{}&
\frac{C_\LF}{\pi  b_0^2}
\biggl\{
\log\!\left(\frac{\as\!\left(\mu_J^2\right)}
{\as\!\left(k_J\mu_K^2/(2 k_K)\right)}\right)
\left[\frac{1}{\as(\theta_{k}^2 k_J^2)} 
- \frac{3 b_0}{4}
\right]
\\&\quad
+ \frac{1}{\as\!\left(\mu_J^2\right)}
- \frac{1}{\as\!\left( k_J\mu_K^2/(2 k_K)\right)}
\biggr\}
\;\;,
\end{split}
\end{equation}
As in the case of gluon splitting, it is possible that, after our approximations, $S_{qqg}$ is negative. Since $S < 0$ is unphysical, we define the complete Sudakov exponent for a quark to be
\begin{equation}
S_q = S_{qqg}\,\Theta(S_{qqg} > 0)
\end{equation}
just to be sure that we are never enhancing an unphysical region by having $e^{-S} > 1$.

Sometimes there is no color connected parton with label $k$ in the fat jet. Then, as in Eq.~(\ref{eq:nokreplacement}) for $S_g$, we make the replacement $\theta_{k} \to R_0$. 

\subsection{After the last splitting}

\begin{figure}
\centerline{\includegraphics[width=10.0cm]{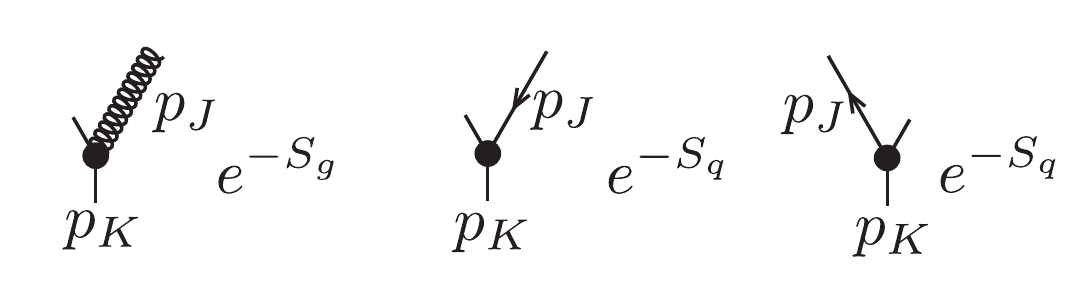}}
\caption{Sudakov factors for partons with no further splittings.}
\label{fig:SudakovEnd}
\end{figure}

If, in the shower history $h$, parton $J$ does not split, then we look at its virtuality $\mu_J^2$ and include a factor $e^{-S_g}$ or $e^{-S_q}$, as illustrated in Fig.~\ref{fig:SudakovEnd}, that represents the probability for parton $J$ not to have split at a virtuality above the final virtuality $\mu_J^2$.

In principle, we should also include a factor $\int\! dH$ representing the probability that parton $J$ did finally split at virtuality $\mu_J^2$. We do not know the splitting angle $\theta$ for this splitting. We do know that $\theta$ was less than $R_{\rm microjet}$, the radius parameter for the $k_T$-jet algorithm that we used to define the microjets: if $\theta$ were larger than $R_{\rm microjet}$, the jet algorithm would not have merged the daughter partons to form the microjet. Thus we would calculate $\int\! dH$ by integrating the differential splitting function over the region $\theta < R_{\rm microjet}$. We do not, in fact, include a splitting factor $\int\! dH$ because this factor is independent of the shower history $h$ and independent of whether we are looking at signal histories or background histories. Thus it cancels from $\chi$. Since we do not need this factor, we do not calculate it.

\subsection{Sudakov factor for initial state emissions}
\label{ISsudakov}

What are the Sudakov factors for the initial state emissions? The initial state emissions can conveniently be ordered according to the value of $k_J^2$. The Sudakov exponent to go from a previous emission scale $k_K^2$ to the new scale $k_J^2$ without a visible initial state emission is, using Eq.~(\ref{eq:ISemission}),
\begin{equation}
\begin{split}
S ={}& \frac{2}{(2\pi)^2}
\int_{k_J^2}^{k_K^2}\! d \bar k^2
\left[\frac{C_\LA}{2}\ 
\frac{\as(\bar k^2 + \kappa_\Lp^2)}{\bar k^2 + \kappa_\Lp^2}\
\frac{1}{(1 + c_R\, \bar k/Q)^{n_R}}
+
\frac{c_{\rm np}(\kappa_{\rm np}^2)^{n_{\rm np} - 1}}
{[\bar k^2 + \kappa_{\rm np}^2]^{n_{\rm np}}}
\right]
\\ & \times
\int\! d\bar y \int\! d\bar \phi\ \Theta(\bar p \in {\rm fat\ jet})
\;\;.
\end{split}
\end{equation}
Here we only count emissions into the region in which the decay products of the emitted parton will be seen as part of the fat jet. Approximately, we can take
\begin{equation}
\int\! d\bar y \int\! d\phi\ \Theta(\bar p \in {\rm fat\ jet})
= \pi R_{\LF}^2
\;\;,
\end{equation}
where $R_\LF$ is the radius parameter that defines the fat jet. Then
\begin{equation}
\begin{split}
S ={}& \frac{R_\LF^2}{2\pi}
\int_{k_J^2}^{k_K^2}\! d \bar k^2
\left[\frac{C_\LA}{2}\ 
\frac{\as(\bar k^2 + \kappa_\Lp^2)}{\bar k^2 + \kappa_\Lp^2}\
\frac{1}{(1 + c_R\, \bar k/Q)^{n_R}}
+
\frac{c_{\rm np}(\kappa_{\rm np}^2)^{n_{\rm np} - 1}}
{[\bar k^2 + \kappa_{\rm np}^2]^{n_{\rm np}}}
\right]
\;\;.
\end{split}
\end{equation}

The initial state shower starts at a transverse momentum scale equal to the scale $Q^2/4$, where $Q^2$ is defined in Eq.~(\ref{eq:Qsqdef}) and represents the scale of the hard interaction. It ends at a scale $k_{\rm cut}^2$, where $k_{\rm cut}$ is the smallest transverse momentum of a microjet that can register in the detector, for instance $k_{\rm cut} = 0.5\ {\rm GeV}$. In general,  there are multiple initial state emissions. We get a Sudakov factor for each one, times a factor for not having an emission between the last one and $k_{\rm cut}^2$. The product of these is $\exp(-S_{\rm IS})$ where
\begin{equation}
\begin{split}
S_{\rm IS} ={}& \frac{R_\LF^2}{2\pi}
\int^{Q^2/4}_{k_{\rm cut}^2}\! d \bar k^2
\left[\frac{C_\LA}{2}\ 
\frac{\as(\bar k^2 + \kappa_\Lp^2)}{\bar k^2 + \kappa_\Lp^2}\
\frac{1}{(1 + c_R\, \bar k/Q)^{n_R}}
+
\frac{c_{\rm np}(\kappa_{\rm np}^2)^{n_{\rm np} - 1}}
{[\bar k^2 + \kappa_{\rm np}^2]^{n_{\rm np}}}
\right]
\;\;.
\end{split}
\end{equation}
The factor $\exp(-S_{\rm IS})$ is independent of the splitting values $k^2_{J_A}$, $k^2_{J_B}$, \dots, $k^2_{J_n}$. It does depend on the hard scattering scale $Q^2$, which varies from event to event. However, note that $Q^2$ is independent of the shower history and is the same for shower histories that represent background and signal processes. Thus the factor $\exp(-S_{\rm tot})$ will cancel exactly between signal and background factors in our observable $\chi$, so we can simply replace
\begin{equation}
\exp(-S_{\rm IS}) \to 1
\;\;.
\end{equation}

\section{Higgs decay probability} 
\label{sec:Higgsdecay}

\begin{figure}
\centerline{\includegraphics[width= 6.0cm]{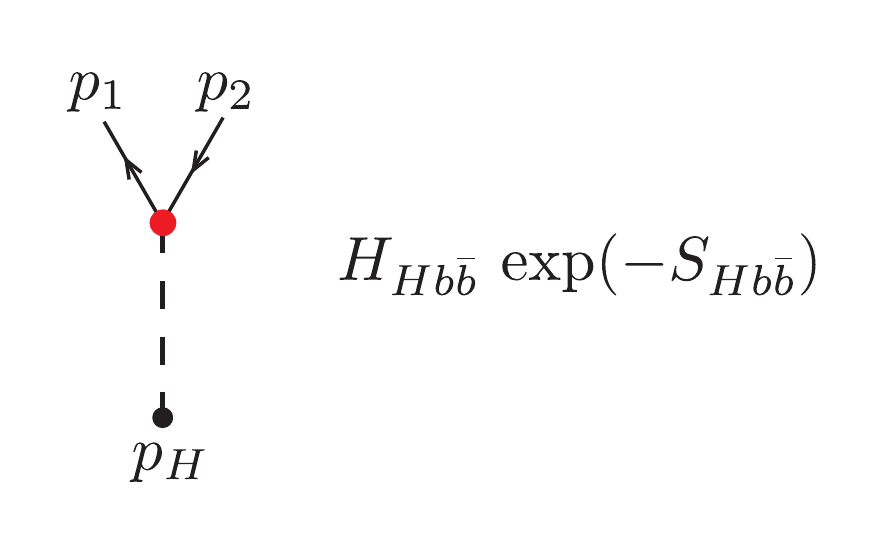}}
\caption{Splitting probability $He^{-S}$ for Higgs boson decay.}
\label{fig:SplittingHbb}
\end{figure}

A light Higgs boson decays most often into $b + \bar{b}$. Since we consider only the $b + \bar{b}$ decay mode, it suffices to treat the Higgs boson as if it always decayed to $b + \bar{b}$. In the sections on splittings in a parton shower, we have specified a conditional splitting probability $H$, the probability for a splitting at a given virtuality $\mu_J^2$ if the parton has not split at a higher $\mu_J^2$. The total splitting probability is then $H e^{-S}$, where $e^{-S}$ is the probability that the parton has not split at a higher $\mu_J^2$. In this section, for the Higgs decay, we specify the total decay probability $H e^{-S}$, depicted in Fig.~\ref{fig:SplittingHbb}.

The light Higgs boson is a very narrow object. In the narrow width approximation, the differential decay probability is
\begin{equation}
H e^{-S} = 16\pi^2\,\delta(m_{b\bar b}^2 - m_H^2)
\;\;.
\end{equation}
The normalization is arranged so that the total probability that the Higgs decays, using the integration measure in in Eq.~(\ref{eq:splitprobability}), is 1:
\begin{equation}
\frac{1}{4(2\pi)^3}\int dm_{b\bar b}^2\int dz\int d\varphi\  H e^{-S} = 1
\;\;.
\end{equation}
%

Although a low mass Higgs boson is a very narrow object, the precision of its mass reconstruction is limited by detector resolution effects and by the loss of momentum resolution caused by grouping final state particles into microjets. To take these issues into account, we treat the Higgs boson decay as if the invariant mass of its decay products can be anything within a $\pm \Delta m_H$ window around the physical Higgs mass, $m_H$. Thus we artificially modify the differential decay probability to
\begin{equation}
\label{eq:Higgsdecay}
H e^{-S} = 16\pi^2\,
\frac{\Theta(|m_{b\bar b} - m_H| < \Delta m_H)}
{4 m_H \,\Delta m_H}
\;\;.
\end{equation}
Our default value for $\Delta m_H$ is 10 GeV.

\section{$b$-tags}
\label{sec:btags}

We have described in Sec.~\ref{sec:FinalStateVariables} how we assign $b$-tags T, F, or {\tt none} to microjets produced by {\sc Pythia} or {\sc Herwig} in a way that mimics imperfect $b$-tagging in an experiment. Tags T or F are assigned only to microjets that are among the three highest $p_T$ microjets in the event and, additionally, have $p_T > p_T^{\rm tag}$, where we take $p_T^{\rm tag} = 15\ {\rm GeV}$.

In this section, we examine how to assign probabilities that a given $b$-tag value will be generated in the simplified shower. We seek to simulate the probabilities with which the algorithm specified above generates $t_j$ values T, F, or {\tt none} when operating on events generated by the full {\sc Pythia} or {\sc Herwig}. 

We suppose that we are given a microjet state, with momenta $p_j$ for each microjet and with a T or F $b$-tag for each microjet that has large enough transverse momentum. We need to estimate the probability $P_j(\LT)$ that microjet $j$ receives a tag $t_{j} = \LT$ and and the probability $P_j(\LF)$ that microjet $j$ receives a tag $t_{j} = \LF$. Then if, in fact,  $t_{j} = \LT$, we include in $P(\{p,t\}_N|\LS,h)$ (for a signal history $h$) or $P(\{p,t\}_N|\LB,h)$ (for a background history $h$) a factor $P_j(\LT)$. If $t_{j} = \LF$, we include factor $P_j(\LF)$. 

How should we calculate $P_j(\LT)$ and $P_j(\LF)$? We note that the situation is simpler than for a real {\sc Pythia} or {\sc Herwig} shower because each microjet consists of precisely one parton and each parton $i$ has a definite flavor $f_i$ which can be $b$ or $\bar b$ or could be a flavor that is not $b$ or $\bar b$, namely $q$ or $\bar q$ or $g$. We make the definition as follows, using the probabilities $P(\LT|b)$ and $P(\LT|\tildenot b)$ defined in Sec.~\ref{sec:FinalStateVariables}:

\begin{quote}

$\bullet$ If a microjet $j$ is a $b$ or $\bar  b$ quark, then we say that $t_{j} = \LT$ with a probability $P_j(\LT) = P(\LT|b)$ and $t_{j} = \LF$ with a probability $P_j(\LF) = 1-P(\LT|b)$.

$\bullet$ If microjet $j$ is not a $b$ or $\bar  b$ quark, then we say that $t_{j} = \LT$ with a probability $P_j(\LT) = P(\LT|\tildenot b)$ and $t_{j} = \LF$ with a probability $P_j(\LF) = 1-P(\LT|\tildenot b)$.

\end{quote}

\section{Constructing shower histories}
\label{ConstructingHistories}

We have now described how to calculate a probability $P(\{p,t\}_N|\LS,h)$ for each signal history $h$ and a probability $P(\{p,t\}_N|\LB,h)$ for each background history $h$. We simply look at the diagram that describes the shower history and associate a factor with each element of the diagram. Now we need to generate shower histories. Because our method for combining daughter jets to form a mother jet is so simple, we can construct a set of possible shower histories in a fairly simple fashion. 

We begin with a list of the starting microjets. We divide these into two sets in all possible ways. One set consists of decay products of partons emitted as initial state or underlying event radiation, the second consists of the decay products of the parton (a gluon for background or a Higgs boson for signal) that is produced in the hard interaction and creates bulk of the fat jet. 

We divide the set of the microjets associated with initial state emissions into any number of non-empty subsets. Each of these subsets is associated with one parton emitted in the initial state.

Now consider the set of microjets associated with the hard parton. In a shower history, the hard parton splits into two partons. The first of these eventually splits to make a subset of the final partons. Call this the set $L$. The second of these eventually splits to make the complementary subset of the final partons. Call this the set $R$. Thus to generate the first splitting of the hard parton, we choose the set $L$ and the set $R$.

For each possible first splitting, we proceed to the second splittings. We can start with the set $L$. We divide this into subsets $LL$ and $LR$. Each of these choices represents a possible splitting. We can simply continue this way until we reach a parton that consists of exactly one microjet.

Each parton emitted in the initial state, as constructed above, consists of a subset of the microjets. If there are more than one microjets in this subset, we can divide it into left and right subsets, which describes a splitting of this parton. Again, this process can be continued until we reach a parton that consists of exactly one microjet.

Note that each parton in the developing shower history consists of a subset of the microjets. Thus we know that the momentum of this parton is $\sum p_i$, summed over this subset. We do not need to know anything about the later shower history of this parton to calculate its momentum. Thus as soon as we have generated a parton splitting, we have the information to calculate the probability for this splitting. The splitting probabilities contain various theta functions that can make the splitting probability equal to zero. When this happens, we can abandon the splitting and try another.

Evidently, the shower histories and the corresponding probabilities can be calculated recursively with a simple computer program. That is what we have done.

\section{Numerical results}
\label{sec:results}

\begin{figure}
\centerline{\includegraphics[width=8.0cm]{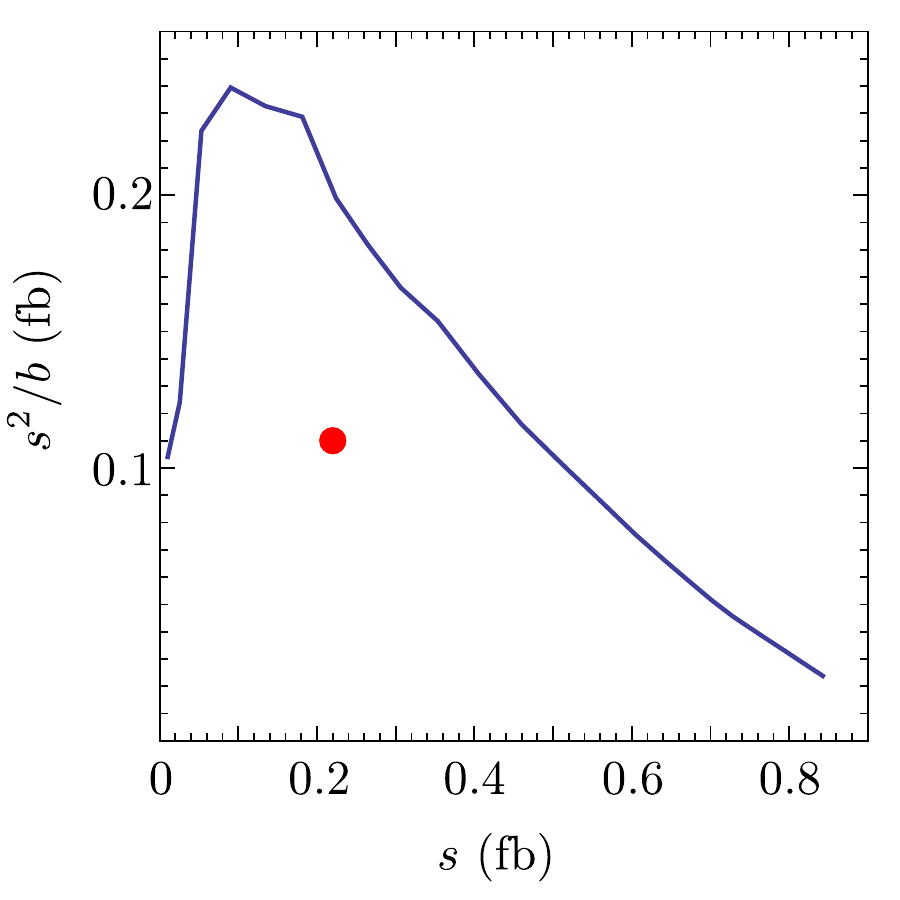}}
\caption{
Plot of $s^2/b$ versus $s$, where $s$ and $b$ are defined in Eq.~(\ref{eq:sandbdef}). We use samples of signal and background events generated by \textsc{Pythia} as in Fig.~\ref{fig:SandBvschi}. This is the same plot as in Fig.~\ref{fig:Rvss} except that we plot $s^2/b$ instead of $s/b$. The total signal cross section with the cuts used is $\sigma_{\rm MC}(\LS) = 1.57\ {\rm fb}$. We also show a point corresponding to a signal cross section $\sigma_{\rm BDRS}(\LS) = 0.22\ {\rm fb}$ and background cross section $\sigma_{\rm BDRS}(\LB) = 0.44\ {\rm fb}$ that we obtained using the method of Ref.~\cite{Butterworth:2008iy}.
}
\label{fig:sRvss}
\end{figure}

We have now seen what shower deconstruction is. In this section, we explore how effective it is for separating signal from background for $p+p \to H + Z + X \to H + \ell^+ + \ell^- + X$. We apply the shower deconstruction method to events generated by {\sc Pythia}, with some comparisons using {\sc Herwig} also. The event selection was described in Sec.~\ref{sec:EventSelection}.

Suppose that we base our analysis on counting events above a cut $\chi$, using the integrated cross sections $s(\chi)$ and $b(\chi)$ defined in Eq.~(\ref{eq:sandbdef}).\footnote{It would be better to use a likelihood ratio based on the full distribution of $ds(\chi)/d\chi$ and $db(\chi)/d\chi$, but the use of a simple cut is easier to describe.} What value of $\chi$ should one choose? If integrated luminosity $\int\! dL$ is available, the expected statistical significance of counting events with $\chi(\{p,t\}_N) > \chi$ is
\begin{equation}
\frac{N(\LS)}{\sqrt{N(\LB)}} = 
\left[
({\textstyle\int\! dL})\
\frac{s(\chi)^2}{b(\chi)}
\right]^{1/2}
\;\;.
\end{equation}
Thus one would choose the value of $\chi$ that maximizes $s^2/b$.

In Fig.~\ref{fig:SandBvschi}, we displayed the $\chi$ distribution for signal and background. We used this information to display $s/b$ as a function of $s$ in Fig.~\ref{fig:Rvss}. In order to understand the statistical significance of a counting experiment with a simple cut on $\chi$, we have seen above that one wants to look at the maximum of $s^2/b$. For that reason, in Fig.~\ref{fig:sRvss}, we display the information from Fig.~\ref{fig:Rvss} as a plot of $s^2/b$ versus $s$.  We have used here the function $\chi(\{p,t\}_N)$ from our simplified shower algorithm. If we could somehow use $\chi_\LMC(\{p,t\}_N)$, using the same Monte Carlo that we use to generate events, we would obtain a curve for $s^2/b$ versus $s$ that is everywhere higher. No algorithm could produce a curve above this limiting curve, but we have no way of determining the limiting curve.

We see in Fig.~\ref{fig:sRvss} that one can achieve a fairly good statistical significance with, say, an integrated luminosity of $\int\! dL = 30\ {\rm fb}^{-1}$. With $s^2/b \approx 0.26$ and this luminosity we have $N(\LS)/\sqrt{N(\LB)} \approx 2.8$. We can compare to the method of Ref.~\cite{Butterworth:2008iy} (BDRS). Applying this method with our data sample, we find a signal cross section $\sigma_{\rm BDRS}(\LS) = 0.22\ {\rm fb}$ and background cross section $\sigma_{\rm BDRS}(\LB) = 0.44\ {\rm fb}$. We have plotted this point in Fig.~\ref{fig:sRvss}. The corresponding statistical significance with $\int\! dL = 30\ {\rm fb}^{-1}$ is 1.8. Of course, this analysis ignores all systematic uncertainties.

In the analysis presented above, we include events with zero, one, and two $b$-tags. Then shower deconstruction has to overcome a signal to background ratio of about 1/1700 in the complete event sample in order to extract a few events with a signal to background ratio of order 1. One suspects that, in fact, the events with zero or one $b$-tags do not contribute much to the discriminating power of the method. Accordingly, we now explore what happens when we give shower deconstruction an easier job by restricting the event sample to just events in which there are two $b$-tagged microjets among the three microjets with the highest transverse momenta that have, additionally, $p_{T} > 15~\rm{GeV}$. With these cuts, the signal sample is 0.39 fb and the background sample is 11 fb. We lose a lot of signal events, but now the signal to background ratio in the event sample is only about 1/30, so the job remaining for shower deconstruction is easier.

\begin{figure}
\centerline{\includegraphics[width=8.0cm]{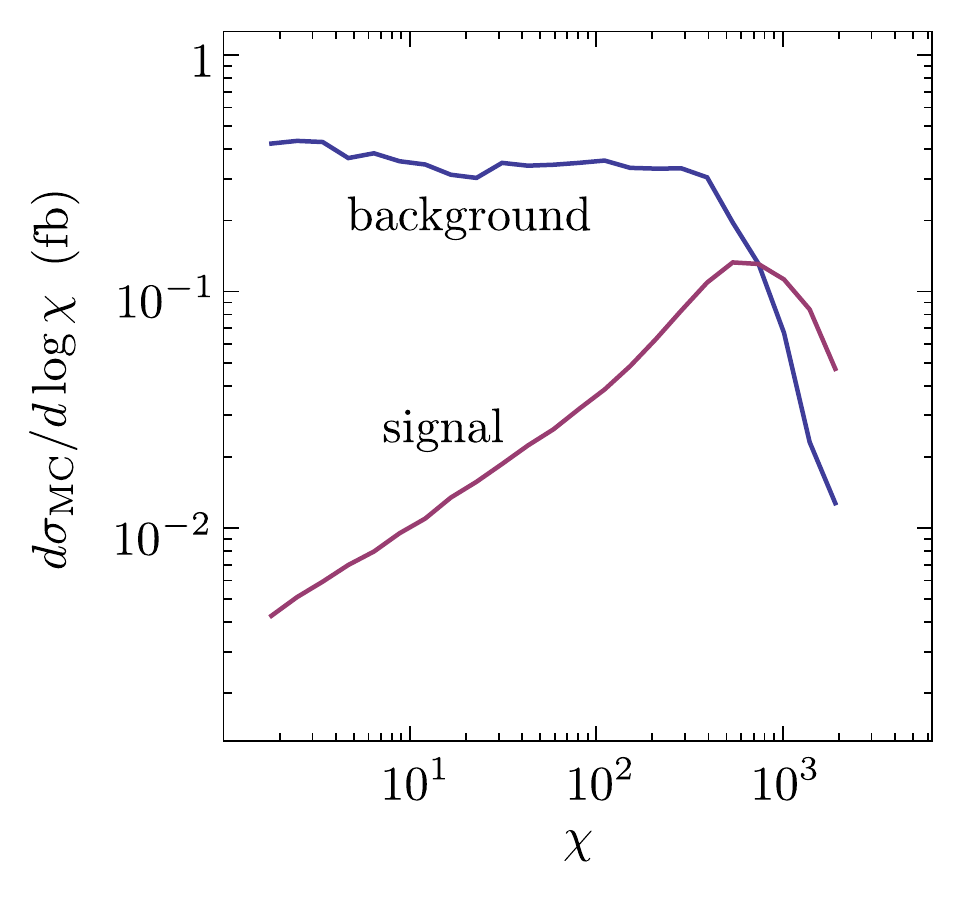}}
\caption{
$d\sigma_\LMC(\LB)/ d\log\chi$ for background events (upper curve) and $d\sigma_\LMC(\LS)/ d\log\chi$ for signal events (lower curve) for samples of signal and background events generated by \textsc{Pythia}. We use the cuts described in Sec.~\ref{sec:EventSelection} and, in addition, require that at least two of the three highest $p_T$ microjets with $p_T > 15\ {\rm GeV}$ have positive $b$-tags.
}
\label{fig:TwobSandBvschi}
\end{figure}

\begin{figure}
\centerline{\includegraphics[width=8.0cm]{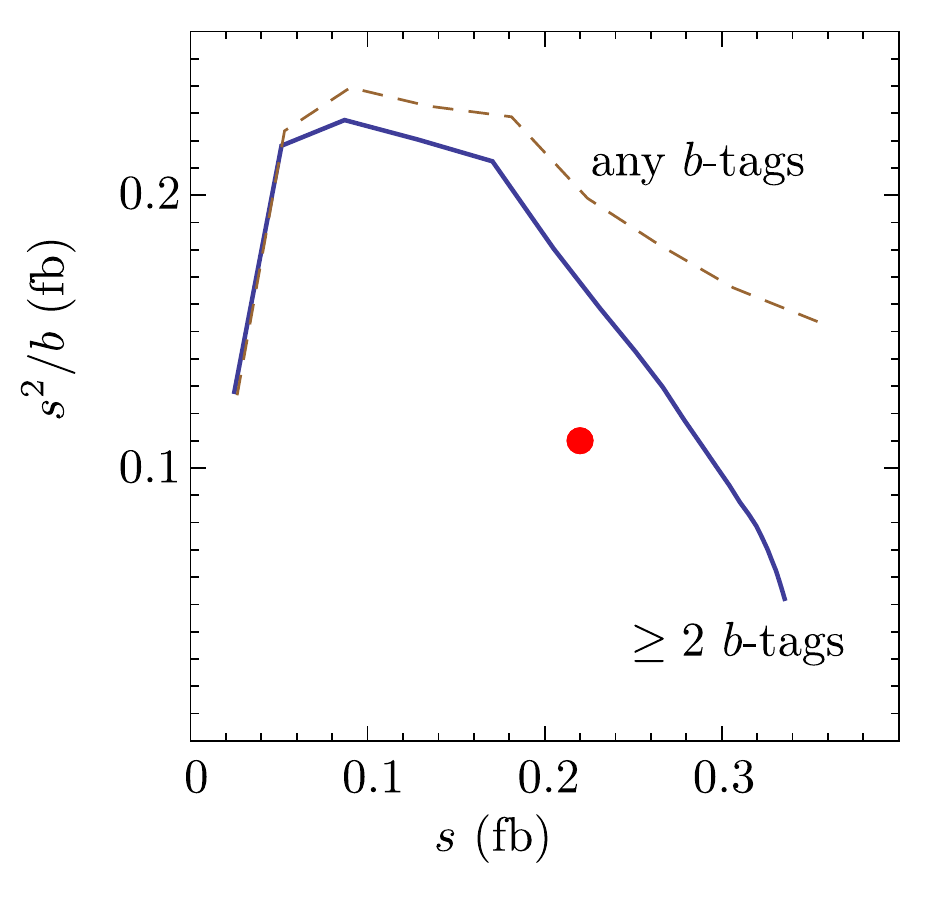}}
\caption{
Plot of $s^2/b$ versus $s$ for events with at least two $b$-tags among the three highest $p_T$ microjets that have $p_{T} > 15~\rm{GeV}$ in addition. We use samples of signal and background events generated by \textsc{Pythia} as in Fig.~\ref{fig:TwobSandBvschi}. We also show the curve from Fig.~\ref{fig:sRvss} for all events with no restriction on $b$-tags (dashed curve) and the point that we obtained using the method of Ref.~\cite{Butterworth:2008iy}.
}
\label{fig:TwobsRvss}
\end{figure}

In Fig.~\ref{fig:TwobSandBvschi} we display the functions $d\sigma_\LMC(\LS)/ d\log\chi$ and $d\sigma_\LMC(\LB)/ d\log\chi$ for the two $b$-tag sample. We again find a region with $s>b$. In Fig.~\ref{fig:TwobsRvss}, we display the information from Fig.~\ref{fig:TwobSandBvschi} as a plot of $s^2/b$ versus $s$. We also show the $s^2/b$ versus $s$ curve from Fig.~\ref{fig:sRvss} for all events with no restriction on $b$-tags and the point that we obtained using the method of Ref.~\cite{Butterworth:2008iy}.\footnote{The method of Ref.~\cite{Butterworth:2008iy} uses only events with two $b$-tags.} We see that for $s \gtrsim 2.5\ {\rm fb}$, $s^2/b$ with the restricted event sample is smaller than it is with the unrestricted event sample. However for  $s \lesssim 2.0\ {\rm fb}$, $s^2/b$ with the restricted event sample is about the same as with the unrestricted event sample.

The formulas that define the simplified shower used to construct Fig.~\ref{fig:TwobsRvss} contain a number of parameters that reflect nonperturbative physics. Among them are $c_{\rm np}$, $\kappa_{\rm np}^2$, $n_{\rm np}$, $c_R$, $n_R$, and $\kappa_\Lp^2$ in Eq.~(\ref{eq:ISemission}), $N_{\rm pdf}^g$ in Eq.~(\ref{eq:Hstart}), and $N_{\rm pdf}^H$ in Eq.~(\ref{eq:signalstart}). There are other parameters like the factor 2 for the hardness cut on splittings in Eq.~(\ref{eq:hardnesscut}) that could have been set differently. We have not systematically tested whether the performance of shower deconstruction as reflected in Fig.~\ref{fig:TwobsRvss} is sensitive to the parameter choices, but we have tried some variations. Typically we found that $d\sigma_\LMC(\LB)/ d\log\chi$ for background events and $d\sigma_\LMC(\LS)/ d\log\chi$ for signal events change in the same direction. Thus we find that the curve in Fig.~\ref{fig:TwobsRvss} is not very sensitive to the parameter variations that we tested.\footnote{We did find that $s^2/b$ could be increased by making the Sudakov exponent for gluon splitting a bit larger, but we have not explored this further.}

\begin{figure}
\centerline{\includegraphics[width=8.0cm]{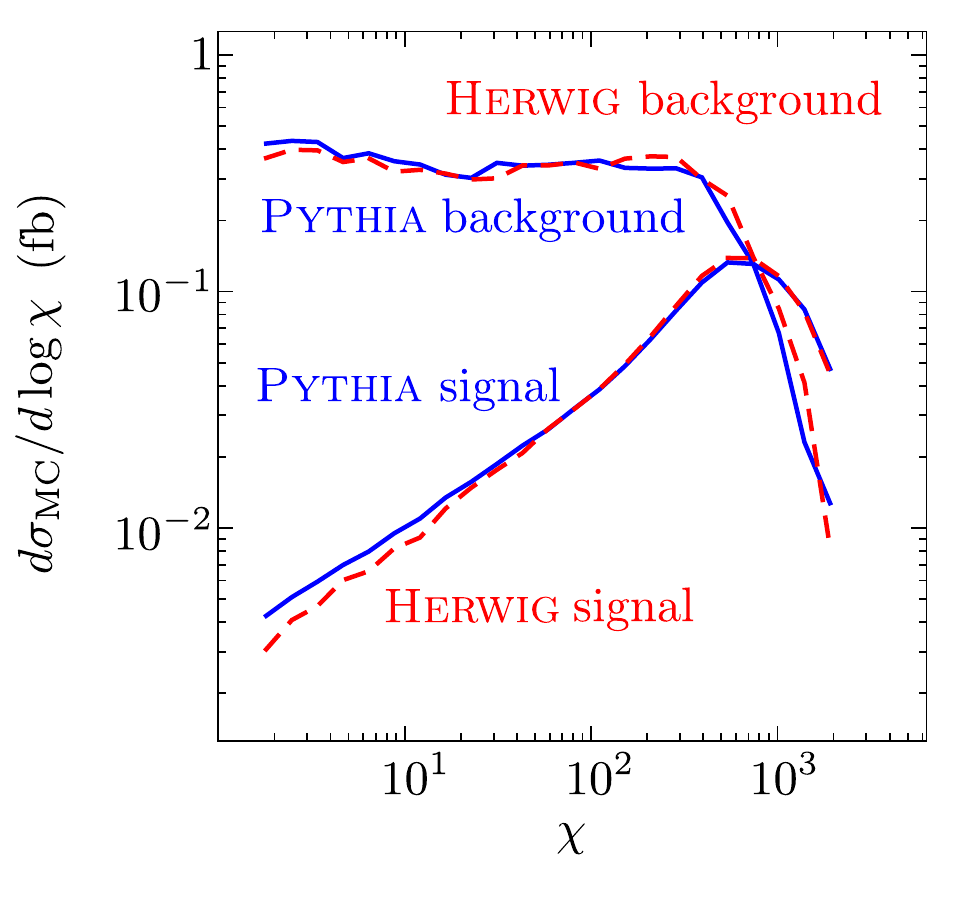}}
\caption{
$d\sigma_\LMC(\LB)/ d\log\chi$ for background events and $d\sigma_\LMC(\LS)/ d\log\chi$ for signal events  for samples of signal and background events generated by \textsc{Pythia} and by \textsc{Herwig}. We use the cuts described in Sec.~\ref{sec:EventSelection} and, in addition, require that at least two of the three highest $p_T$ microjets with $p_T > 15\ {\rm GeV}$ have positive $b$-tags. The solid (blue) lines are for {\sc Pythia} while the dashed (red) lines are for {\sc Herwig}. At small $\chi$, the background curves are on the top and the signal curves are on the bottom.
}
\label{fig:HPvschi}
\end{figure}

We have used {\sc Pythia} \cite{Pythia} for our comparisons. What would happen if we used {\sc Herwig} \cite{Herwig} instead? We show in Fig.~\ref{fig:HPvschi} the cross sections $d\sigma_\LMC(\LB)/ d\log\chi$ and $d\sigma_\LMC(\LS)/ d\log\chi$ for two $b$-tag samples of signal and background events generated by \textsc{Pythia} and by \textsc{Herwig}. We have normalized the cross sections within our cuts to be the same for both \textsc{Pythia} and \textsc{Herwig}, so that we are looking at differences in shape rather than normalization. We see that the behaviors obtained with the two event generators are quite similar but that with \textsc{Herwig} a somewhat larger fraction of the background events have large $\chi$. That there are differences is not a surprise since both event generators work at leading order in perturbation theory for their splitting kernels and make approximations with respect to color and spin of partons. One lesson from this is that in experimental applications of shower deconstruction or of other jet substructure measures one will want to test the Monte Carlo cross sections against experiment.

\begin{figure}
\centerline{\includegraphics[width=8.0cm]{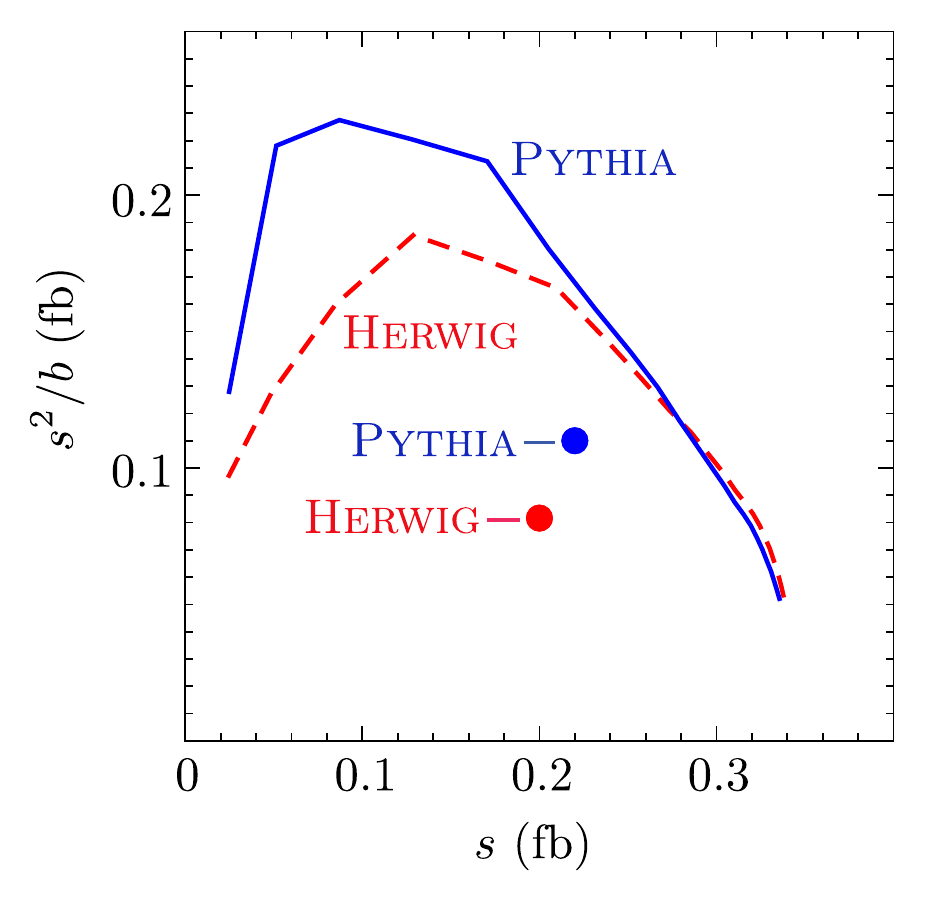}}
\caption{
Plot of $s^2/b$ versus $s$ for events with two positive $b$-tags. We compare the distribution of $s^2/b$ for events generated with {\sc Pythia}
as in Fig.~\ref{fig:TwobsRvss}, to the same distribution using events generated with {\sc Herwig}. We normalize the total signal and background cross sections with these cuts to be $\sigma_{\rm MC}(\LS) = 0.39\ {\rm fb}$, $\sigma_{\rm MC}(\LB) = 11\ {\rm fb}$.  We also show points that we obtained using the method of Ref.~\cite{Butterworth:2008iy}. Using {\sc Pythia} we found $\sigma_{\rm BDRS}(\LS) = 0.22\ {\rm fb}$ and $\sigma_{\rm BDRS}(\LB) = 0.44\ {\rm fb}$, as in Fig.~\ref{fig:TwobsRvss}, while using {\sc Herwig} we found $\sigma_{\rm BDRS}(\LS) = 0.20\ {\rm fb}$ and $\sigma_{\rm BDRS}(\LB) = 0.49\ {\rm fb}$.
}
\label{fig:PythiaHerwig}
\end{figure}

In Fig.~\ref{fig:PythiaHerwig} we compare results from the two $b$-tag sample using \textsc{Pythia} and \textsc{Herwig} for $s^2/b$ as a function of $s$. We also show results using \textsc{Pythia} and \textsc{Herwig} for $s^2/b$ using the BDRS method. For \textsc{Pythia}, these are the results that were exhibited in Fig.~\ref{fig:TwobsRvss}. We see that there is about a 30\% difference between \textsc{Pythia} and \textsc{Herwig} results. Again, this level of difference using leading order event generators is not a surprise. 

\section{Conclusions}
\label{sec:conclusions}

We have proposed a method, shower deconstruction, for separating signal and background events when we have a definite theory in mind for the signal as well as for the standard model background with the signal process omitted. We have explained the method using a simple signal process, $p+p \to H + Z + X \to H + \ell^+ + \ell^- + X$. Here the event selection is chosen so that the Higgs boson that we hope to find is boosted to a substantial transverse momentum. The shower deconstruction method itself is quite general and could be applied to signal processes with more structure or perhaps to signal processes in which the sought massive objects are not highly boosted.

The idea of shower deconstruction can be described in very few words. With data at hand, one begins by clustering final state particles in a region of the detector (the ``fat jet'' in our example) into much smaller jets, the microjets, using the $k_T$-jet algorithm. Alternatively one could use some other jet algorithm or one could use topological clusters defined directly using the calorimetry of the experiment. This gives a fairly fine grained description of the event, with the momenta $p_i$ and possibly flavor tags $t_i$ for each microjet. In order to keep within reasonable bounds for computer resources, one can limit the number $N$ of microjets by discarding the lowest transverse momentum microjets as necessary. One wants to be fine grained enough to see not only the direct decay products of a sought heavy particle but also gluon radiation that reflects the color structure of the signal or background final state. Then one computes approximately the probability $P(\{p,t\}_N|\LS)$ to obtain the observed microjet state $\{p,t\}_N$ from the signal process and the probability $P(\{p,t\}_N|\LB)$ to obtain the microjet state from a background process. We construct the observable $\chi(\{p,t\}_N) = {P(\{p,t\}_N|\LS)}/{P(\{p,t\}_N|\LB)}$ as the ratio of these and use $\chi$ to distinguish signal from background. The value of $\chi$ is calculated using a simplified shower algorithm that tries to mimic what a partitioned dipole shower with initial state radiation and underlying event contributions would give. The microjets are treated as intermediate state partons in the shower. We want the calculation to be as accurate as possible, but it needs to be an analytic calculation that can be executed with a not-to-large amount of computer time for each event. There is a tension between these goals. We expect that other workers will be able to improve on the compromise algorithm that we have described in this paper.

This method is similar in spirit to the matrix element method \cite{Kondo:1988yd,Kondo:1991dw,Fiedler:2010sg,Alwall:2010cq}. There, if one started from the microjet configuration $\{p,t\}_N$, one would compute $\chi(\{p,t\}_N)$ from the squared matrix element for the signal or background process convoluted with the parton distribution functions, integrated over the momenta of unobserved partons. If one were to use a number of partons $N$ that is greater than the minimum possible number for the desired signal and background and if one were to calculate $\chi(\{p,t\}_N)$ analytically, one would have something close to the shower deconstruction method. In one sense, one would then have a better approximation to nature than the simplified shower algorithm of this paper because one would be using the exact squared matrix element rather than a soft-collinear approximation to it. However, one would be missing the Sudakov factors. Without Sudakov factors, the probability for a parton splitting becomes infinite as the virtuality of the splitting tends to zero. With Sudakov factors, the probability for a parton splitting approaches zero as the virtuality of the splitting tends to zero. For this reason, one needs the Sudakov factors.

We have found that in our simple example the shower deconstruction can achieve a signal/background discrimination superior to that of Ref.~\cite{Butterworth:2008iy}. Furthermore, shower deconstruction has some features that suggest that it may prove useful as a practical tool. First, it is quite general, although further development is needed to apply the general method to other signal processes. Second, it is modular, with modules corresponding to QCD parton splitting, initial state radiation, underlying event contributions, Sudakov factors, and heavy particle decay. The modules can be improved independently and inserted into the general scheme. Third, the method has at least the potential to work for quite complicated signal processes.

\acknowledgments{ 
This work was supported by the United States Department of Energy. Its genesis was at the workshop on jets and jet substructure at the LHC that took place at the University of Washington in January 2010, with Department of Energy sponsorship. We thank Steve Ellis for organizing that workshop. Most of what author DS knows about parton showers has arisen from working with Zoltan Nagy; the present paper is an application of that knowledge. We also thank Zoltan Nagy and David Reeb for conversations about the specifics of this paper and we thank Gustaaf Brooijmans and Peter Loch for advice about calorimetry. We thank Daniel Steck and Jeremy Thorn for the use of the Quantum Control computer cluster at the University of Oregon.
}
\appendix
\section{The jacobian}
\label{sec:jacobiancalc}

In this appendix, we analyze the integral
\begin{equation}
\begin{split}
\label{eq:IJdefagain}
I_J \equiv {}&
\frac{1}{16}
\int\!dk_A^2
\int\!dy_A
\int\!d\phi_A 
\int\!dk_B^2
\int\!dy_B
\int\!d\phi_B\
\delta^4(p_A + p_B - p_J)
\times f
\;\;.
\end{split}
\end{equation}
Here $p_A$ and $p_B$ are the momenta of two jets that together form the jet with momentum $p_J$. In our application, the two constituent jets have non-zero masses, $\mu_A$ and $\mu_B$. However, the masses $\mu_A$ and $\mu_B$ are small compared to the jet transverse momenta $k_A$ and $k_B$ and compared to the combined jet mass, $\mu_J$. Thus it is a good approximation to neglect the constituent jet masses; furthermore, doing so leads to a substantially simpler result. We therefore set $\mu_A = \mu_B = 0$. With this choice, the $(+,-,1,2)$ components of the momenta of the jets are (with $p^\pm = (p^0 \pm p^3)/\sqrt 2$)
\begin{equation}
\begin{split}
\label{eq:momentumdecompositions}
p_A ={}& \left(
\frac{1}{\sqrt 2}\,k_A\, e^{y_A},
\frac{1}{\sqrt 2}\,k_A\, e^{-y_A},
k_A \cos\phi_A,
k_A \sin\phi_A
\right)
\;\;,
\\
p_B ={}& \left(
\frac{1}{\sqrt 2}\,k_B\, e^{y_B},
\frac{1}{\sqrt 2}\,k_B, e^{-y_B},
k_B \cos\phi_B,
k_B \sin\phi_B
\right)
\;\;,
\\
p_J ={}& \left(
\frac{1}{\sqrt 2}\,\sqrt{k_J^2 + \mu_J^2}\, e^{y_J},
\frac{1}{\sqrt 2}\,\sqrt{k_J^2 + \mu_J^2}\, e^{-y_J},
k_J \cos\phi_J,
k_J \sin\phi_J
\right)
\;\;.
\end{split}
\end{equation}

We wish to write $I_J$ in the form
\begin{equation}
\label{eq:Jdef}
I_J = \int\!dz \int\!d\varphi\ J \times f
\;\;.
\end{equation}
Here $z$ is a momentum fraction defined by
\begin{equation}
\label{eq:zdef}
z = \frac{k_A}{k_A + k_B}
\;\;.
\end{equation}
Then
\begin{equation}
1-z = \frac{k_B}{k_A + k_B}
\end{equation}
and
\begin{equation}
z(1-z) = \frac{k_A k_B}{(k_A + k_B)^2}
\;\;.
\end{equation}
We define the variable $\varphi$ by
\begin{equation}
\label{eq:tanvarphi}
\tan\varphi =
\frac{\sinh(\Delta y/2)\cos(\Delta\phi/2)}{\cosh(\Delta y/2)\sin(\Delta\phi/2)}
\;\;,
\end{equation}
where
\begin{equation}
\begin{split}
\Delta y ={}& y_A - y_B
\;\;,
\\
\Delta \phi ={}& \phi_A - \phi_B
\;\;.
\end{split}
\end{equation}
Thus $\varphi$ is approximately the angle about the origin in the $(\Delta \phi,\Delta y)$ plane. We need to calculate the jacobian $J$.

To proceed, we define unit vectors
\begin{equation}
\begin{split}
n_0 ={}& \left(\frac{1}{\sqrt 2}\,e^{y_J}, \frac{1}{\sqrt 2}\,e^{-y_J},0,0\right)
\;\;,
\\
n_3 ={}& \left(-\frac{1}{\sqrt 2}\,e^{y_J}, \frac{1}{\sqrt 2}\,e^{-y_J},0,0\right)
\;\;,
\\
n_1 ={}& \left(0, 0, \cos\phi_J, \sin\phi_J\right)
\;\;,
\\
n_2 ={}& \left(0, 0, -\sin\phi_J, \cos\phi_J\right)
\;\;.
\end{split}
\end{equation}
These are orthogonal to each other and normalized as unit vectors along the coordinate axes in a convenient reference frame: $n_\mu \cdot n_\nu = g_{\mu\nu}$. We thus have
\begin{equation}
\begin{split}
\label{eq:IJdef1}
I_J = {}&
\frac{1}{16}
\int\!dy_A
\int\!dy_B
\int\!dk_A^2
\int\!dk_B^2
\int\!d\phi_A
\int\!d\phi_B
\\ &\times
\delta((p_A + p_B - p_J)\cdot n_1)\,
\delta((p_A + p_B - p_J)\cdot n_2)
\\ &\times
\delta((p_A + p_B - p_J)\cdot n_0)\,
\delta((p_A + p_B - p_J)\cdot n_3)
\times f
\;\;.
\end{split}
\end{equation}

Let us examine the effect of
\begin{equation}
\delta((p_A + p_B - p_J)\cdot n_2)
= \delta((\bm k_A + \bm k_B)\cdot \bm n_2)
= \delta(k_A\sin(\phi_A - \phi_J) + k_B\sin(\phi_B - \phi_J))
\;\;.
\end{equation}
Here we use boldface symbols to represent transverse vectors. We can use the delta function to perform the integration over $\phi_B$:
\begin{equation}
\int\!d\phi_A \int\!d\phi_B\
\delta((p_A + p_B - p_J)\cdot n_2)
\cdots
= \int\!d\phi_A\ \frac{1}{k_B\,|\cos(\phi_B - \phi_J)|}
\cdots
\;\;.
\end{equation}
Then
\begin{equation}
\label{eq:getphiB}
\sin(\phi_B - \phi_J) = - \frac{k_A}{k_B}\sin(\phi_A - \phi_J)
\;\;.
\end{equation}
We want to change the integration variable to $\Delta \phi = \phi_A - \phi_B$. From Eq.~(\ref{eq:getphiB}) we have
\begin{equation}
k_A\sin(\phi_A-\phi_J) = - k_B[
\cos(\phi_A - \phi_B)\sin(\phi_A - \phi_J)
-\sin(\phi_A - \phi_B)\cos(\phi_A - \phi_J)
]
\;\;.
\end{equation}
That is
\begin{equation}
\tan(\phi_A-\phi_J) = \frac{k_B \sin \Delta \phi}{k_A + k_B \cos \Delta \phi}
\;\;.
\end{equation}
Thus
\begin{equation}
d(\phi_A-\phi_J)
= k_B \cos^2(\phi_A-\phi_J)\,
\frac{k_B + k_A \cos \Delta \phi}{(k_A + k_B \cos \Delta \phi)^2}\ 
d\Delta\phi
\;\;.
\end{equation}
We also derive
\begin{equation}
\cos^2(\phi_A - \phi_J) = 
\frac{(k_A + k_B \cos\Delta \phi)^2}{k_A^2 + k_B^2 +  2 k_A k_B \cos\Delta \phi}
\;\;,
\end{equation}
so
\begin{equation}
d(\phi_A-\phi_J)
= k_B  
\frac{k_B + k_A \cos \Delta \phi}
{k_A^2 + k_B^2 +  2 k_A k_B \cos\Delta \phi}\ d\Delta \phi
\;\;.
\end{equation}
Since also
\begin{equation}
\cos^2(\phi_B - \phi_J) = 
\frac{(k_B + k_A \cos\Delta \phi)^2}{k_A^2 + k_B^2 +  2 k_A k_B \cos\Delta \phi}
\;\;,
\end{equation}
we have
\begin{equation}
d(\phi_A-\phi_J)
=   
\frac{k_B\cos(\phi_B - \phi_J)}
{[k_A^2 + k_B^2 +  2 k_A k_B \cos\Delta \phi]^{1/2}}\ d\Delta \phi
\;\;.
\end{equation}
Additionally, we note that
\begin{equation}
[k_A^2 + k_B^2 +  2 k_A k_B \cos\Delta \phi]^{1/2} = k_J
\;\;.
\end{equation}
Thus
\begin{equation}
\int\!d\phi_A\int\!d\phi_B\
\delta((p_A + p_B - p_J)\cdot n_2)
\cdots
= \int\!\frac{d\Delta \phi}
{k_J}
\cdots
\;\;.
\end{equation}
With this result, we have
\begin{equation}
\begin{split}
\label{IJ1}
I_J = {}&
\frac{1}{16}
\int\!dy_A
\int\!dy_B
\int\!dk_A^2
\int\!dk_B^2
\int\!\frac{d\Delta \phi}
{k_J}
\\ &\times
\delta((p_A + p_B - p_J)\cdot n_1)\,
\delta((p_A + p_B - p_J)\cdot n_0)\,
\delta((p_A + p_B - p_J)\cdot n_3)
\times f
\;\;.
\end{split}
\end{equation}

We next turn to the elimination of the delta function with $n_3$. We note that
\begin{equation}
\delta((p_A + p_B - p_J)\cdot n_3)
=
\delta(k_A\sinh(y_A-y_J) + k_B\sinh(y_B-y_J))
\;\;.
\end{equation}
We can use this delta function to eliminate the integration over $y_B$:
\begin{equation}
\int\!dy_A\int\!dy_B\
\delta((p_A + p_B - p_J)\cdot n_3)
\cdots
= \int\!dy_A\ \frac{1}{k_B\,\cosh(y_B-y_J)}
\cdots
\;\;.
\end{equation}

We want to change the integration variable to $\Delta y = y_A - y_B$. We have
\begin{equation}
k_A\sinh(y_A-y_J) = - k_B\sinh(y_B-y_J)
\;\;.
\end{equation}
Thus
\begin{equation}
k_A\sinh(y_A-y_J) = - k_B[
\cosh(y_A - y_B)\sinh(y_A - y_J)
-\sinh(y_A - y_B)\cosh(y_A - y_J)
]
\;\;.
\end{equation}
That is
\begin{equation}
\tanh(y_A-y_J) = \frac{k_B \sinh \Delta y}{k_A + k_B \cosh \Delta y}
\;\;.
\end{equation}
Thus
\begin{equation}
d(y_A-y_J)
= k_B \cosh^2(y_A-y_J)\,
\frac{k_B + k_A \cosh \Delta y}{(k_A + k_B \cosh \Delta y)^2}\ d\Delta y
\;\;.
\end{equation}
We also derive
\begin{equation}
\cosh^2(y_A - y_J) = 
\frac{(k_A + k_B \cosh\Delta y)^2}{k_A^2 + k_B^2 +  2 k_A k_B \cosh\Delta y}
\;\;,
\end{equation}
so
\begin{equation}
d(y_A-y_J)
= k_B  
\frac{k_B + k_A \cosh \Delta y}
{k_A^2 + k_B^2 +  2 k_A k_B \cosh\Delta y}\ d\Delta y
\;\;.
\end{equation}
Since also
\begin{equation}
\cosh^2(y_B - y_J) = 
\frac{(k_B + k_A \cosh\Delta y)^2}{k_A^2 + k_B^2 +  2 k_A k_B \cosh\Delta y}
\;\;,
\end{equation}
we have
\begin{equation}
d(y_A-y_J)
=   
\frac{k_B\cosh(y_B - y_J)}
{[k_A^2 + k_B^2 +  2 k_A k_B \cosh\Delta y]^{1/2}}\ d\Delta y
\;\;.
\end{equation}
We also note that
\begin{equation}
\begin{split}
k_A^2 + k_B^2 +  2 k_A k_B \cosh\Delta y ={}& 
k_A^2  + k_B^2   + 2 k_A k_B\cos\Delta\phi
\\&
+ 2 k_A k_B (\cosh\Delta y
- \cos\Delta\phi)
\\
={}& k_J^2 + 2 p_A\cdot p_B
\\
={}& k_J^2 + \mu_J^2
\;\;.
\end{split}
\end{equation}
Thus
\begin{equation}
\int\!dy_A\int\!dy_B\
\delta((p_A + p_B - p_J)\cdot n_3)
\cdots
= \int\!\frac{d\Delta y}
{\sqrt{k_J^2 + \mu_J^2}}
\cdots
\;\;.
\end{equation}
With this result, we have
\begin{equation}
\begin{split}
\label{eq:IJ2}
I_J = {}&
\frac{1}{16}
\int\!dk_A^2
\int\!dk_B^2
\int\!\frac{d\Delta \phi}
{k_J}
\int\!\frac{d\Delta y}
{\sqrt{k_J^2 + \mu_J^2}}
\\ &\times
\delta((p_A + p_B - p_J)\cdot n_1)\,
\delta((p_A + p_B - p_J)\cdot n_0)
\times f
\;\;.
\end{split}
\end{equation}
Now we would like to use the remaining delta functions to eliminate the integrations over $k_A^2$ and $k_B^2$. 

For the delta function involving $n_0$, we have
\begin{equation}
\delta((p_A + p_B - p_J)\cdot n_0)
=
\delta\!\left(k_A\cosh(y_A-y_J) + k_B\cosh(y_B-y_J) - a_J
\right)
\;\;,
\end{equation}
where we abbreviate
\begin{equation}
a_J = \sqrt{k_J^2 + \mu_J^2}
\;\;.
\end{equation}
Using our results expressing $\cosh(y_A-y_J)$ and $\cosh(y_B-y_J)$ in terms of $\Delta y$, this is
\begin{equation}
\delta((p_A + p_B - p_J)\cdot n_0)
=
\delta\!\left(
\frac{k_A(k_A + k_B \cosh\Delta y) + k_B(k_B + k_A \cosh\Delta y)}
{[k_A^2 + k_B^2 +  2 k_A k_B \cosh\Delta y]^{1/2}}
- a_J
\right)
\;\;.
\end{equation}
That is
\begin{equation}
\delta((p_A + p_B - p_J)\cdot n_0)
=
\delta\left(
[k_A^2 + k_B^2 +  2 k_A k_B \cosh\Delta y]^{1/2}
- a_J
\right)
\;\;.
\end{equation}
We can write
\begin{equation}
[k_A^2 + k_B^2 +  2 k_A k_B \cosh\Delta y]^{1/2} - a_J
=\frac{k_A^2 + k_B^2 +  2 k_A k_B \cosh\Delta y - a_J^2}
{[k_A^2 + k_B^2 +  2 k_A k_B \cosh\Delta y]^{1/2} + a_J}
\;\;.
\end{equation}
The denominator is not singular, so we can factor it out and evaluate it at the point at which the numerator vanishes:
\begin{equation}
\label{eq:deltan0}
\delta((p_A + p_B - p_J)\cdot n_0)
=
2 a_J\,
\delta\left(
k_A^2 + k_B^2 +  2 k_A k_B \cosh\Delta y
- a_J^2
\right)
\;\;.
\end{equation}
We will use this result below at Eq.~(\ref{eq:IJ3}).

For the delta function involving $n_1$, we have
\begin{equation}
\delta((p_A + p_B - p_J)\cdot n_1)
=
\delta( k_A\cos(\phi_A-\phi_J) + k_B\cos(\phi_B-\phi_J) - k_J)
\;\;.
\end{equation}
Using our results expressing $\cos(\phi_A-\phi_J)$ and $\cos(\phi_B-\phi_J)$ in terms of $\Delta \phi$, this is
\begin{equation}
\delta((p_A + p_B - p_J)\cdot n_1)
=
\delta\!\left(
\frac{k_A(k_A + k_B \cos\Delta \phi) + k_B(k_B + k_A \cos\Delta \phi)}
{[k_A^2 + k_B^2 +  2 k_A k_B \cos\Delta \phi]^{1/2}}
- k_J
\right)
\;\;.
\end{equation}
That is
\begin{equation}
\delta((p_A + p_B - p_J)\cdot n_1)
=
\delta\left(
[k_A^2 + k_B^2 +  2 k_A k_B \cos\Delta \phi]^{1/2}
- k_J
\right)
\;\;.
\end{equation}
We can write
\begin{equation}
[k_A^2 + k_B^2 +  2 k_A k_B \cos\Delta\phi]^{1/2} - k_J
=\frac{k_A^2 + k_B^2 +  2 k_A k_B \cos\Delta\phi - k_J^2}
{[k_A^2 + k_B^2 +  2 k_A k_B \cos\Delta\phi]^{1/2} + k_J}
\;\;.
\end{equation}
The denominator is not singular, so we can factor it out and evaluate it at the point at which the numerator vanishes:
\begin{equation}
\delta((p_A + p_B - p_J)\cdot n_1)
=
2 k_J\,
\delta\!\left(
k_A^2 + k_B^2 +  2 k_A k_B \cos\Delta\phi
- k_J^2
\right)
\;\;.
\end{equation}
It will prove convenient to write this as
\begin{equation}
\label{eq:n1result}
\delta((p_A + p_B - p_J)\cdot n_1)
=
2 k_J\,
\delta\!\left(
(k_A  + k_B)^2 -  2 k_A k_B (1 - \cos\Delta\phi)
- k_J^2
\right)
\;\;.
\end{equation}
We will use this result below at Eq.~(\ref{eq:dkAdkB}).

Now let us change integration variables to $(k_A + k_B)^2$ and $2 k_A k_B$, with
\begin{equation}
dk_A^2\,dk_B^2 = \frac{k_A k_B}{|k_A^2 - k_B^2|}\ d(k_A + k_B)^2\, d(2 k_A k_B)
\;\;.
\end{equation}
When we make this change of variables, we ought to introduce also a sum over the discrete variable that distinguishes between $k_A$ and $k_B$, since $(k_A + k_B)$ and $(2 k_A k_B)$ are invariant under interchange of $k_A$ and $k_B$. However, we omit a special notation for this because we will soon change back to a variable $z$ that does distinguish between $k_A$ and $k_B$.

We can eliminate the integration over $(k_A + k_B)^2$ at fixed $2 k_A k_B$ using the $n_1$ delta function from Eq.~(\ref{eq:n1result}):
\begin{equation}
\label{eq:dkAdkB}
dk_A^2\,dk_B^2\ \delta((p_A + p_B - p_J)\cdot n_1) 
= d(2 k_A k_B)\ \frac{2 k_J k_A k_B}{|k_A^2 - k_B^2|}
\;\;.
\end{equation}
Here
\begin{equation}
\label{eq:kApluskB}
(k_A  + k_B)^2 =  k_J^2 + 2 k_A k_B (1 - \cos\Delta\phi)
\;\;.
\end{equation}

This gives
\begin{equation}
\begin{split}
\label{eq:IJ3}
I_J = {}&
\frac{1}{8}
\int\!d\Delta \phi
\int\!d\Delta y
\int\!dt\ \frac{t}{|k_A^2 - k_B^2|}\,
\delta(A(t))
\times f
\;\;,
\end{split}
\end{equation}
where we have defined
\begin{equation}
t = 2 k_A k_B
\end{equation}
and where $A$ is the argument of the delta function in Eq.~(\ref{eq:deltan0}),
\begin{equation}
A
=
k_A^2 + k_B^2 
+ 2 k_A k_B \cosh\Delta y
- k_J^2 - \mu_J^2
\;\;.
\end{equation}
From Eq.~(\ref{eq:kApluskB}), we have
\begin{equation}
k_A^2 + k_B^2 = k_J^2 - t \cos\Delta\phi
\;\;.
\end{equation}
Thus
\begin{equation}
A(t)
=
t\, (\cosh\Delta y - \cos\Delta\phi)
- \mu_J^2
\;\;,
\end{equation}
so that
\begin{equation}
\begin{split}
\label{eq:IJ4}
I_J = {}&
\frac{1}{4}
\int\!d\Delta \phi
\int\!d\Delta y\
\frac{k_A k_B}{|k_A^2 - k_B^2|}\,
\frac{1}{\cosh\Delta y - \cos\Delta\phi}
\times f
\;\;,
\end{split}
\end{equation}
where
\begin{equation}
\label{eq:tresult}
2k_A k_B\, (\cosh\Delta y - \cos\Delta\phi)
= \mu_J^2
\;\;.
\end{equation}

This nearly completes the task set at the beginning of this appendix. Now, let us change to some more useful integration variables. 

Let us define a momentum fraction $z$ according to Eq.~(\ref{eq:zdef}). We need to express $z(1-z)$ as a function of $\Delta y$ and $\Delta \phi$. Using Eqs.~(\ref{eq:tresult}) and (\ref{eq:kApluskB}), we have
\begin{equation}
\begin{split}
k_A k_B ={}& \frac{\mu_J^2/2}{\cosh\Delta y - \cos\Delta\phi}
\;\;,
\\
(k_A  + k_B)^2 ={}&  \frac{k_J^2(\cosh\Delta y - \cos\Delta\phi) 
+ \mu_J^2(1 - \cos\Delta\phi)}
{\cosh\Delta y - \cos\Delta\phi}
\;\;.
\end{split}
\end{equation}
Thus
\begin{equation}
\label{eq:z1mz}
z(1-z) = \frac{\mu_J^2/2}
{k_J^2(\cosh\Delta y - \cos\Delta\phi) 
+ \mu_J^2(1 - \cos\Delta\phi)}
\;\;.
\end{equation}
From this, we calculate
\begin{equation}
\begin{split}
\frac{\partial z(1-z)}{\partial \Delta\phi} ={}& 
- \frac{2z^2(1-z)^2k_J^2}{\mu_J^2}\,(1+R)
\sin \Delta\phi
\;\;,
\\
\frac{\partial z(1-z)}{\partial \Delta y} ={}& 
- \frac{2z^2(1-z)^2 k_J^2 }{\mu_J^2}\,\sinh \Delta y
\;\;.
\end{split}
\end{equation}
where
\begin{equation}
R = \frac{\mu_J^2}{k_J^2}
\;\;.
\end{equation}

We need another variable, $\varphi$, which we define according to Eq.~(\ref{eq:tanvarphi}). The gradient of $\tan\varphi$ is
\begin{equation}
\begin{split}
\frac{\partial \tan\varphi}{\partial \Delta\phi} ={}& 
- \frac{\tan\varphi}{\sin \Delta\phi}
\;\;,
\\
\frac{\partial \tan\varphi}{\partial \Delta y} ={}& 
\frac{\tan\varphi }{\sinh \Delta y}
\;\;.
\end{split}
\end{equation}
We can use the partial derivatives to calculate the jacobian, giving
\begin{equation}
d\Delta\phi\ d\Delta y =
\frac{\mu_J^2}{2 z^2(1-z)^2 k_J^2}\
\frac{\sinh\Delta y\,  \sin\Delta\phi}
{\sinh^2\Delta y + (1+R)\sin^2\Delta\phi}\
d(z(1-z))\ \frac{d\tan\varphi}{\tan\varphi}
\;\;.
\end{equation}
That is,
\begin{equation}
\label{eq:phiytozvarphi}
d\Delta\phi\ d\Delta y =
\frac{\mu_J^2}{2 z^2(1-z)^2 k_J^2 }\
\frac{|1-2z|}
{\sinh^2\Delta y + (1+R)\sin^2\Delta\phi}\
\frac{\sinh\Delta y\,  \sin\Delta\phi}{\sin\varphi \cos\varphi}\
dz\ d\varphi
\end{equation}
With a little algebra, we find
\begin{equation}
\frac{1}{\sin\varphi \cos\varphi}
= 2\frac{\cosh \Delta y - \cos \Delta\phi}{\sinh\Delta y\,\sin\Delta\phi}
\;\;.
\end{equation}
Thus
\begin{equation}
\label{eq:phiytozvarphi2}
d\Delta\phi\ d\Delta y =
\frac{\mu_J^2}{z^2(1-z)^2 k_J^2 }\
\frac{|1-2z| [\cosh \Delta y - \cos \Delta\phi]}
{\sinh^2\Delta y + (1+R)\sin^2\Delta\phi}\
dz\ d\varphi
\end{equation}

Now we insert this result into Eq.~(\ref{eq:IJ4}). There is a factor
\begin{equation}
\frac{k_A k_B}{|k_A^2 - k_B^2|}
= \frac{z(1-z)}{|1-2z|}
\;\;,
\end{equation}
which cancels the $|1-2z|$ in the numerator of Eq.~(\ref{eq:phiytozvarphi}). Then
\begin{equation}
\begin{split}
\label{eq:IJ5}
I_J = {}&
\frac{1}{4}
\int\!dz
\int\!d\varphi\
\frac{\mu_J^2}{z(1-z) k_J^2 }\
\frac{1}
{\sinh^2\Delta y + (1+R)\sin^2\Delta\phi}\
\times f
\;\;.
\end{split}
\end{equation}
We can use Eq.~(\ref{eq:z1mz}) to express $\mu_J^2$ in terms of $z(1-z)$ and the angles $(\Delta y,\Delta\phi)$:
\begin{equation}
\begin{split}
\label{eq:IJ6}
I_J = {}&
\frac{1}{2}
\int\!dz
\int\!d\varphi\
\frac{(\cosh\Delta y - \cos\Delta\phi) 
+ R(1 - \cos\Delta\phi)}
{\sinh^2\Delta y + (1+R)\sin^2\Delta\phi}\
\times f
\;\;.
\end{split}
\end{equation}
We can rewrite this as
\begin{equation}
\begin{split}
\label{eq:IJ7}
I_J = {}&
\frac{1}{4}
\int\!dz
\int\!d\varphi\
\frac{\sinh^2(\Delta y/2) + (1+R)\sin^2(\Delta\phi/2)}
{\sinh^2(\Delta y/2)\cosh^2(\Delta y/2) 
+ (1+R)\sin^2(\Delta\phi/2)\cos^2(\Delta\phi/2)}\
\times f
\;\;.
\end{split}
\end{equation}

This is the result that we sought. We note that since $\cosh(\Delta y/2) \approx 1$ and $\cos(\Delta\phi/2) \approx 1$ for small angles $(\Delta y,\Delta \phi)$, we have approximately
\begin{equation}
\begin{split}
\label{eq:IJ8}
I_J \approx {}&
\frac{1}{4}
\int\!dz
\int\!d\varphi\ f
\;\;.
\end{split}
\end{equation}
when the integration is dominated by the small angle region.


\end{document}